\documentclass[aps,onecolumn,nofootinbib,superscriptaddress,letterpaper,amsmath,amssymb]{revtex4}
\pdfoutput=1

\usepackage{booktabs}
\usepackage{graphicx}  
\usepackage{dcolumn}   
\usepackage{bbm}        
\usepackage{amsmath}
\usepackage{amsfonts}
\usepackage{amssymb}   
\usepackage{epstopdf}
\usepackage{slashed}
\usepackage{hyperref}

\usepackage{multirow}
\usepackage{color}
\usepackage{float}
\usepackage{ragged2e}
\usepackage{subcaption}
\usepackage[switch]{lineno}
\DeclareCaptionJustification{justified}{\justifying}
\usepackage[normalem]{ulem}
\usepackage{verbatim}
\usepackage{xspace}

\def\gsim{\lower0.5ex\hbox{$\:\buildrel >\over\sim\:$}}
\def\lsim{\lower0.5ex\hbox{$\:\buildrel <\over\sim\:$}}

\newcommand{\secref}[1]{{Sec.~\ref{#1}}}

\definecolor{ccorange}{rgb}{1.0,0.5,0.0}
\newcommand{\sjangle}{\psi}   
\newcommand{\pt}{$p_{\rm T}$}

\newcommand{\be}{\begin{equation}}
\newcommand{\ee}{\end{equation}}
\newcommand{\bea}{\begin{eqnarray}}
\newcommand{\eea}{\end{eqnarray}}

\newcommand{\nbox}{{\,\lower0.9pt\vbox{\hrule \hbox{\vrule height 0.2 cm
\hskip 0.2 cm \vrule height 0.2 cm}\hrule}\,}}

\def\to{\rightarrow}

\newskip\zatskip \zatskip=0pt plus0pt minus0pt
\def\matth{\mathsurround=0pt}
\def\lsim{\mathrel{\mathpalette\atversim<}}
\def\gsim{\mathrel{\mathpalette\atversim>}}
\def\sigv{\ifmmode \langle\sigma v\rangle\else $\langle\sigma v\rangle$\fi}
\newskip\zatskip \zatskip=0pt plus0pt minus0pt
\def\matth{\mathsurround=0pt}
\def\lsim{\mathrel{\mathpalette\atversim<}}
\def\gsim{\mathrel{\mathpalette\atversim>}}
\def\atversim#1#2{\lower0.7ex\vbox{\baselineskip\zatskip\lineskip\zatskip
  \lineskiplimit
  0pt\ialign{$\matth#1\hfil##\hfil$\crcr#2\crcr\sim\crcr}}}

\begin{document}
\thispagestyle{empty}

\vspace{0.5in}

\title{Learning to Calibrate Jet Energies}
\author{Bogdan Malaescu}
\affiliation{LPNHE, Sorbonne Universit\'e, Universit\'e Paris Cit\'e, CNRS/IN2P3, Paris, France}
\author{Benjamin Nachman}
\affiliation{Department of Particle Physics and Astrophysics, Stanford University, Stanford, CA 94305, USA}
\affiliation{Fundamental Physics Directorate, SLAC National Accelerator Laboratory, Menlo Park, CA 94025, USA}
\author{Mukul Ram}
\affiliation{Independent Researcher}
\author{Steven Schramm}
\affiliation{D\'epartement de Physique Nucl\' eaire et Corpusculaire, Universit\'e de Gen\'eve, Gen\`eve;
Switzerland}
\author{Daniel Whiteson}
\affiliation{Department of Physics \& Astronomy, University of California, Irvine, CA}




\begin{abstract}

Hadronic jets are produced copiously at the Large Hadron Collider, and are a vital element of its physics program. Accurate reconstruction of their energies is therefore crucial, including correction for pile-up, detector response and data-simulation differences. Currently, each correction is performed by fitting
parametric functional forms, an approach which is effective but can be fragile and challenging to extend. We present a prototype machine-learning pipeline which performs pile-up dependency removal, absolute corrections and data-mc corrections. No functional form is imposed on the correction, additional features can be easily added, and the pipeline matches or exceeds the performance of the functional approach at nearly every stage.
\end{abstract}

\maketitle

\section{Introduction}
\label{Sec:Introduction}

Hadronic jets, collimated sprays of particles, are produced copiously in the proton--proton collisions of the
Large Hadron Collider, often dominating the decay channels of massive particles~\cite{ATLAS:2018kot,CMS:2018nsn} and the most common background processes~\cite{ATLAS:2019fgd,CMS:2019gwf}.  
A vast program of measurements and searches therefore rely on accurate reconstruction and reliable assessment of associated systematic uncertainties for their energies and directions~\cite{ATLAS:2014rjc,CMS:2016lmd}.  
Reconstruction of the true jet energy from the detector responses is non-trivial, requiring accounting for effects including contributions from additional $pp$ interactions, varying detector response and discrepancies between simulation and data~\cite{ATLAS:2015ull,CMS:2016lmd}.  

The current calibration schemes used by ATLAS~\cite{ATLAS:2014rjc,ATLAS:2019oxp,ATLAS:2020cli,ATLAS:2018bip} and CMS~\cite{CMS:2011shu,Agarwal:2022txa} apply several sequential corrections by extracting parameters from fits of
parametric functional forms to selected observables. Such corrections are well motivated physically but are insensitive to dependence on other features, impose simplifying assumptions and are increasingly difficult to scale to larger datasets and additional observables.  Jets whose response depends on observables the
parametrization cannot accommodate are mis-calibrated, and the error is
absorbed into the systematic uncertainty rather than corrected.

Machine learning methods offer greater performance and flexibility than fits to parametric forms~\cite{Guest:2018yhq}. Indeed, neural networks have already demonstrated the capacity to correct the jet response in a way that is sensitive to the jet's internal composition, at equal or better
resolution~\cite{ATLAS:2023hbp,ATLAS:2020nnc} than traditional methods. Networks operating on jet constituents have also been used
to predict the particle-level energy directly~\cite{CMS:2019bjr,ATLAS:2023dnn}. Neural networks have also been used to parametrize detector response~\cite{DetectorResponseNN}. Beyond direct regression, whose result depends on the spectrum of the training sample~\cite{MLCalibrationBias}, neural networks have been used for numerical inversion~\cite{ATLAS:2018gni,ATLAS:2020nnc}, discriminative maximum-likelihood calibration~\cite{FrequentistCalibration} and calibration with conditional generative models~\cite{CaloFlowCalibration}. Machine-learning methods have also been developed for pile-up mitigation~\cite{PUMML,PileupGNN}. These efforts generally target individual stages of reconstruction or calibration; comparatively little attention has been paid to a machine-learning treatment of the complete calibration chain, including pile-up dependency removal and residual data--simulation corrections. A complete machine-learning-based pipeline offers greater extendability and lays the groundwork for a simplified, single-step calibration.

\begin{figure}
\includegraphics[width=0.95\linewidth]{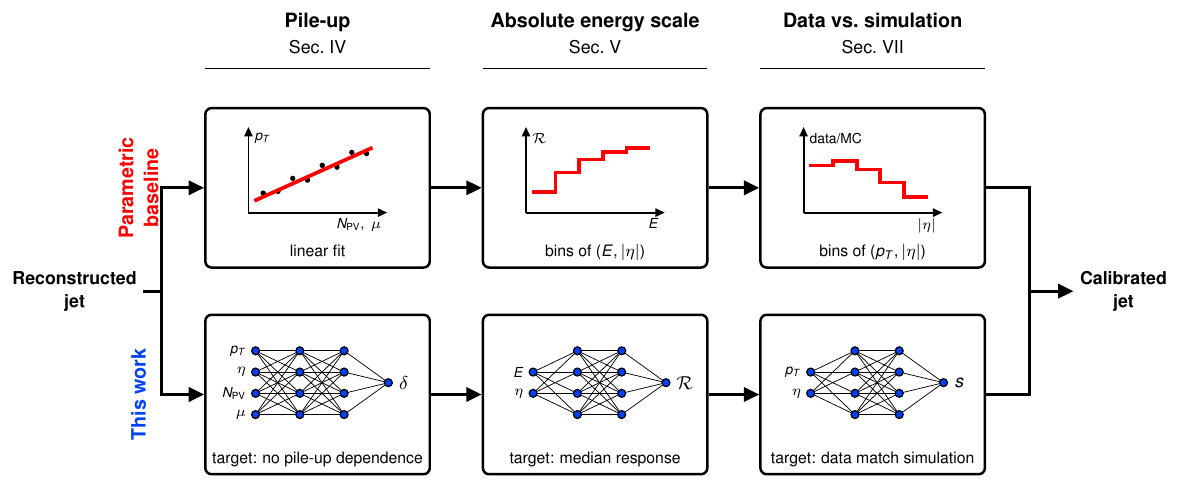}
    \caption{Schematic view of the two pipelines that calibrate the energy of reconstructed jets: the neural-network-based approach described here, and the parametric approach, which follows ATLAS, used as a baseline.}
    \label{fig:schem}
\end{figure}

Here, we present a prototype of a machine-learning-based pipeline in which every
stage is performed by a neural network trained directly against the quantity that
stage is meant to correct; see Fig~\ref{fig:schem}.  No functional form is imposed on the correction, and additional
features can be simply added without engineering new parametric forms to fight the curse of dimensionality; we demonstrate
this by adding the positions of the two leading sub-jets to the calibration of
large-radius jets which straddle calorimeter cracks, where the response varies
rapidly and depends on where the sub-jets fall. The pipeline matches or exceeds
the performance of the current approach at nearly every stage and is  compatible with  nuisance parameter decomposition and reduction~\cite{ATLAS:2014rjc}.

This paper is organized as follows. Section~\ref{sec:background} defines the
calibration task and describes the baseline procedure against which we compare.
Section~\ref{sec:samples} describes the simulated samples and the distortions
injected to make the task realistic. Section~\ref{sec:pileup} treats the
subtraction of pile-up and Section~\ref{sec:abs} the correction of the average
energy response. Section\ref{sec:subjet_y} extends that
correction to large-radius jets using substructure information.
Section~\ref{sec:mcdata} treats the correction of residual differences between
data and simulation, and Section~\ref{sec:np} the nuisance parameters which carry
its uncertainty. Section~\ref{sec:discussion} discusses the results and the work
remaining.

\section{Background}
\label{sec:background}

\subsection{Jets}

Jets are collimated sprays of hadrons produced by the showering and hadronization of quarks and gluons. The
\emph{parton} is the unobservable quark or gluon. The
\emph{particle-level jet} is a cluster of stable particles, mostly hadrons, produced by the parton, grouped by a jet-finding algorithm and only available in simulated samples. The
\emph{reconstructed jet} is a cluster of detector deposits grouped by a jet-finding algorithm.

The calibration seeks to correct the energy of the reconstructed jet, the measured object, to the particle-level jet, the object a perfect detector would measure, where any difference represents  experimental mismeasurement or resolution.  The
ratio of the two energies is the \emph{response}, $\mathcal{R} = E^{\rm reco} /
E^{\rm true}$, and a perfectly calibrated jet is one whose response is unity. The difference between the particle-level jet
and the parton, in contrast, depends on showering and hadronization theories as well as the jet algorithm.

In this paper, we focus on three major effects which separate the reconstructed and particle-level jet energies. Additional $pp$ collisions in the same and nearby
bunch crossings, collectively \emph{pile-up}, deposit energy in the region the jet
occupies and add to its measured momentum; a {\it pile-up correction} addresses this. The calorimeter registers less energy
from a hadronic shower than from an electromagnetic one of the same energy, and
its response varies further with energy, with pseudorapidity and across the
transitions between its components; an {\it absolute correction} corrects for this. Finally, the calibration is performed in simulation, which may not describe the physical detector exactly, leading to mis-calibration; an {\it in-situ correction} moves the data towards the simulation.  

\subsection{\texorpdfstring{Parametric baseline}{Parametric baseline}}

The three stages of correction as implemented by the ATLAS collaboration~\cite{ATLAS:2020cli} are roughly reproduced here as a baseline, and described below. The implementation follows ATLAS for concreteness; since each of its stages fits a fixed functional form, it is called the {\it parametric baseline} below.

The pile-up correction is performed in two steps, first removing the expected contribution and then subtracting remaining residual dependence. The median transverse-momentum density in the event is measured, $\rho$, and the product of the density and the jet area $A$ is removed~\cite{CacciariSalamPileup,ATLAS:2015ull}.  The second step measures the residual pileup contribution's dependence on the number of reconstructed vertices ($N_{\textrm{PV}}$) and on the mean number
of interactions per crossing ($\mu$), and fits a linear form in several bins of pseudorapidity and transverse momentum. The applied correction is taken as the value of the linear coefficients  at $p_{\textrm{T}}^{\rm true} = 25$~GeV, so the applied
correction carries no dependence on transverse momentum. Full details are given in Sec.~\ref{sec:pileup_atlas}.

The absolute jet energy scale is set in simulated samples, using matched pairs of particle-level and reconstructed jets  to calculate the  \emph{response}, $\mathcal{R} = E^{\rm reco} /
E^{\rm true}$. The response is profiled in bins of
true energy and pseudorapidity, smoothed, and inverted numerically so that a
correction derived against the true energy can be applied to the measured one~\cite{Cukierman:2016ntt,ATLAS:2018gni}.
Details are given in Sec.~\ref{sec:abs_atlas}. A further stage, the global sequential
calibration, improves the resolution using observables describing the internal
composition of the jet without changing the average response. It is not modelled
here, since neural networks have already been shown to reproduce
it in prior work~\cite{ATLAS:2023hbp} by other authors.

Residual differences between simulation and data are removed last, by balancing
jets against objects whose momenta are better measured: dijet events fix the
response of forward jets relative to central ones, a $Z$ boson or photon fixes the
central scale in absolute terms, and a system of calibrated jets extends the range
beyond the reach of either boson, or more recently the balance of energy and momentum~\cite{ATLAS:2023hbp}. Further details are given in Sec.~\ref{sec:mcdata_atlas}.

The CMS calibration~\cite{CMS:2016lmd} has the same sequence of stages. An offset correction (L1) removes the pile-up energy using $\rho$ and the jet area; a correction derived from simulation (L2L3) brings the response to the particle level as a function of transverse momentum and pseudorapidity; and residual corrections derived in data, relative (L2Res) from dijet events and absolute (L3Res) from $Z$+jet, $\gamma$+jet and multijet events, balance the jet both directly and through the projection of the missing transverse momentum onto the reference (MPF). The approach described here replaces the stages one by one and applies equally to either chain.

\section{Samples}
\label{sec:samples}

Two simulated samples are used, a sample of QCD dijet events which supports the bulk
of the study and a sample of large-radius jets from hadronic W decays used for the substructure
extension of \secref{sec:subjet_y}. In both, the calibration target is the
particle-level jet rather than the initiating parton.

\subsection{Dijet sample}

\begin{figure}
    \centering
    \includegraphics[width=0.3\linewidth]{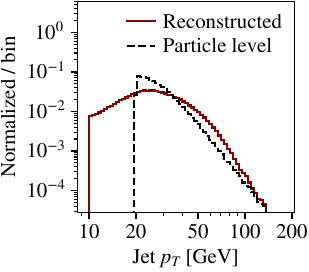}
        \includegraphics[width=0.3\linewidth]{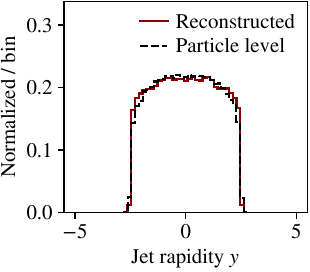}
        \includegraphics[width=0.3\linewidth]{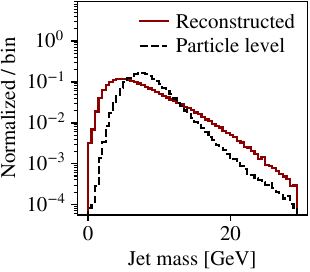}
    \caption{ Distributions of jet transverse momentum, rapidity and mass for particle-level and reconstructed jets in a sample of simulated $pp\rightarrow jj$ collisions.}
    \label{fig:jet_kin}
\end{figure}

Dijet events, $pp\rightarrow jj$, $j\in [u,d,s,c,b,g]$ are generated with Madgraph5~\cite{madgraph} at LO, with \textsc{Pythia}~8~\cite{Pythia8}
for the parton shower and hadronization, with the parton transverse momentum
required to exceed 20~GeV. The detector response is simulated with
\textsc{Delphes}~\cite{delphes} using its parameterization of the ATLAS detector.
Reconstructed jets are clustered with the anti-$k_t$ algorithm~\cite{Cacciari:2008gp} with
radius parameter $R=0.5$ from calorimeter towers, the energy deposits in the electromagnetic and hadronic calorimeter cells, each kept only if its energy exceeds a noise threshold; they play the role of the topological clusters of ATLAS~\cite{ATLASTopoClusters}. Particle-level jets are built by applying the same algorithm to the stable
generated particles. The sample is defined at particle level: each particle-level jet with $p_{\textrm{T}}>20$~GeV is matched to the nearest reconstructed jet with $|\eta|<2.5$ within $\Delta R=\sqrt{\Delta y^{2}+\Delta\phi^{2}} < 0.3$, the separation in rapidity and azimuth, provided that jet is not nearer to another particle-level jet. Reconstructed jets are kept down to 10~GeV. These matched pairs are the unit on which 
corrections are derived and evaluated. Example kinematic distributions are
shown in Fig.~\ref{fig:jet_kin}.

Two quantities required by the area-based pile-up subtraction are recorded during
reconstruction. Each jet carries its area, the region of rapidity and azimuth
over which it collects energy. Each event carries $\rho$, the median transverse
momentum per unit area of the calorimeter towers, taken over a grid of $0.5\times0.5$ cells in rapidity and azimuth within $|y|<2.5$.

\begin{figure}
    \centering
    \includegraphics[width=0.3\linewidth]{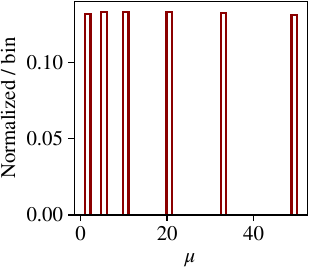}
        \includegraphics[width=0.3\linewidth]{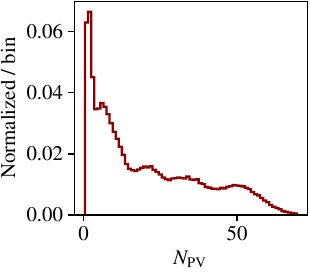}
        \includegraphics[width=0.3\linewidth]{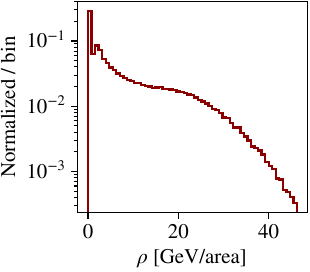}
    \caption{ Distributions of the mean number of interactions ($\mu$), the number of reconstructed primary vertices ($N_{PV}$) and the median transverse momentum per unit area in a sample of simulated $pp\rightarrow jj$ collisions.}

    \label{fig:jet_pu}
\end{figure}

Pile-up is simulated by overlaying additional $pp$ interactions at six values of
the mean number of interactions per bunch crossing, $\mu\in\{1,5,10,20,33,50\}$.
The resulting distributions of $\mu$, of the number of reconstructed primary
vertices $N_{\rm PV}$ and of $\rho$ over all events in all samples are shown in Fig.~\ref{fig:jet_pu}.
\textsc{Delphes} assigns a single value of $\mu$ to each sample rather than
drawing it per event and performs no vertex reconstruction, so $N_{\rm PV}$ and
$\mu$ carry less independent information here than in data. The consequences for
the pile-up correction are treated in \secref{sec:pileup}.

\begin{figure}
    \centering
    \includegraphics[width=0.3\linewidth]{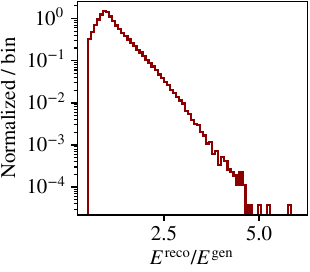}
        \includegraphics[width=0.3\linewidth]{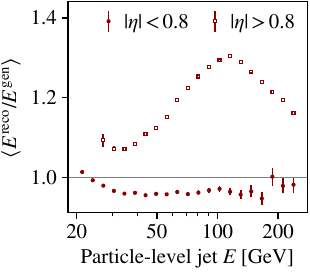}
         \includegraphics[width=0.3\linewidth]{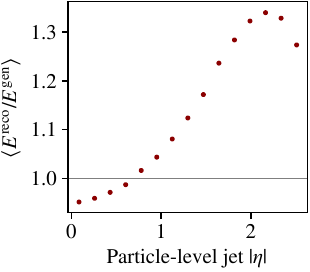}
    \caption{ Distributions of the  detector response ($E^{reco}/E^{gen}$) and mean response versus particle-level energy, for central and forward jets, and pseudorapidity in a sample of simulated $pp\rightarrow jj$ collisions described in the text.}
    \label{fig:jet_res}
\end{figure}

The response of the simulated detector is shown in Fig.~\ref{fig:jet_res}, both
inclusively and as the mean response as a function of particle-level energy and
pseudorapidity.   Before pile-up is subtracted, as shown there, the mean response is close to unity for central jets and rises with $|\eta|$ because at the same transverse momentum, more energy is deposited above the calorimeter noise thresholds in the forward region.  \textsc{Delphes} reproduces the average behavior of a calorimeter but not its detailed structure: there are no transitions between calorimeter regions and no intrinsic difference between the response to electromagnetic and hadronic showers.

\subsection{Large-radius jet sample}

The substructure extension in Sec~\ref{sec:subjet_y} requires jets whose energy is shared between two
resolved sub-jets. Events containing a photon and a hadronically decaying $W$
boson, $pp\rightarrow\gamma W$ with $W\rightarrow q\bar q$, are generated with
\textsc{MadGraph}~\cite{madgraph} for the hard process and showered as above. The photon
is required to have transverse momentum above 225~GeV, so that the $W$ recoils
against it with enough momentum for its two quarks to merge into a single jet.
Pile-up is overlaid as in the dijet sample.

Jets are clustered with the anti-$k_t$ algorithm with radius parameter $R=1.0$ from the same calorimeter towers. No grooming is applied to the jets: the energy that is distorted and calibrated in
\secref{sec:subjet_y} is that of the ungroomed anti-$k_t$ jet. The soft-drop
algorithm~\cite{Larkoski:2014wba,Dasgupta:2013ihk}, with $\beta = 0$ and
$z_{\mathrm{cut}} = 0.1$, is used only to identify the two sub-jets: the jet is
declustered step by step and the softer branch is discarded until both branches carry
at least a fraction $z_{\mathrm{cut}}$ of their combined transverse momentum, and those
two branches are the sub-jets. They are retained for each jet at both reconstructed and
particle level, and their positions are the additional features supplied to the network
in \secref{sec:subjet_y}. Particle-level jets with $|\eta|<2.5$, transverse momentum above 225~GeV and two soft-drop sub-jets are matched one-to-one to reconstructed jets with $|\eta|<2.5$ and two soft-drop sub-jets, within $\Delta R<0.4$.

\subsection{Injected distortions}

The detector effects \textsc{Delphes} models are simpler in structure
than those of a real experiment. Additional distortions are
therefore injected by hand, each reproducing a feature of the real calibration
problem that the fast simulation omits, and each applied before the stage that is
meant to remove it. They are described where they are used, the pile-up
dependence in \secref{sec:pileup}, the energy-scale distortion in
\secref{sec:abs} and the data-simulation differences in \secref{sec:mcdata}.

\section{Pile-up}
\label{sec:pileup}

Energy deposited by additional $pp$ interactions, {\it pile-up}, is
the first contribution the calibration removes. This section describes the
pile-up dependence injected into the samples, the implementation of the
parametric correction as a benchmark, our NN-based method, 
the metrics used to judge both approaches and the results. A second injection, in which the
dependence arises from a mis-estimated pile-up density rather than being added
directly, is described in \secref{sec:etadens}.

\subsection{Injection of pileup dependence}

The first step in pile-up mitigation is area-based subtraction~\cite{CacciariSalamPileup}, in which
the product of the event pile-up density $\rho$ and the jet area $A$ is removed
from the jet transverse momentum. What remains is the residual, the difference
between the reconstructed and the particle-level transverse momentum, and the
dependence of that residual on the pile-up conditions is what this stage of the
calibration aims to remove.

\begin{figure}
    \centering
    \includegraphics[width=0.95\linewidth]{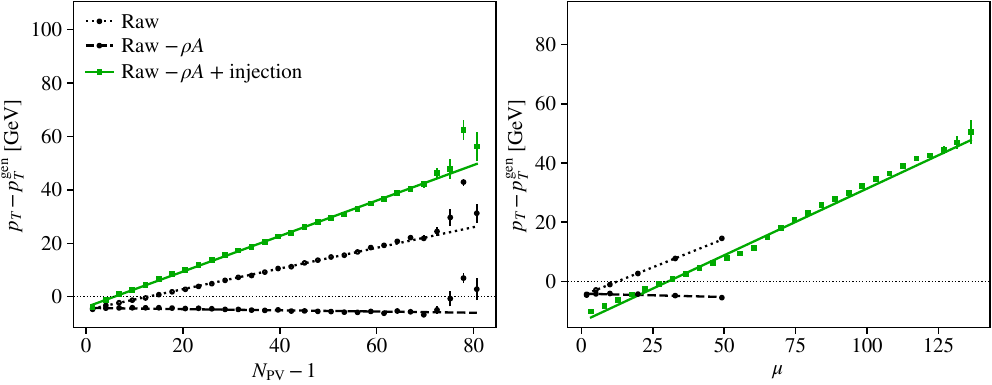}
    \caption{ Dependence of the residual between true and generation jet transverse momentum on the number of primary vertices, $N_{PV}$ and on the mean number of interactions, $\mu$. Shown are the mean residuals in the simulated samples before $\rho A$ correction, after $\rho A$ correction and after our injection described in the text.}
    \label{fig:pu_inject}
\end{figure}

That dependence is seen in ATLAS's  samples with realistic simulation, but nearly absent in the simulated sample of $pp\rightarrow jj$ collisions described above. Figure~\ref{fig:pu_inject} shows the residual against $N_{\rm PV}$ and against
$\mu$, before and after the subtraction of $\rho \times A$. The strong dependence
present beforehand is significantly suppressed, leaving little for a subsequent
correction to act on. This weak dependence is due to \textsc{Delphes}' relatively simple description of  pile-up, assigning one
value of $\mu$ to each sample and reconstructing no vertices, such that simulated
pile-up is more uniform and more completely described by $\rho$ than  pile-up
in a real experiment.  Additional dependence is therefore injected.

\begin{figure}
    \centering
    \includegraphics[width=0.32\linewidth]{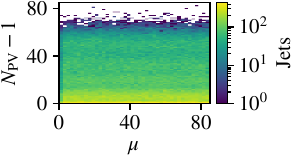}
        \includegraphics[width=0.32\linewidth]{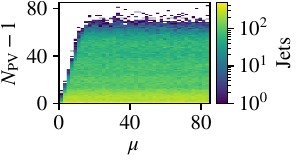}
    \includegraphics[width=0.32\linewidth]{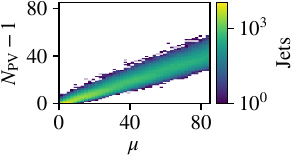}
    \caption{ Examples of correlation between two observables which are sensitive to instantaneous luminosity, the mean number of interactions ($\mu$) and the number of reconstructed primary vertices ($N_{PV}$). Three examples of correlation are shown:  $\rho_{\mu N} =$ 0 (left), 0.1 (center), 1.0 (right). }
    \label{fig:pu_corr_sim}
\end{figure}

The two variables which describe the pile-up conditions, $N_{\rm PV}$ and $\mu$,
both track the instantaneous luminosity and are therefore correlated in data, so
dependence assigned to one of them is not cleanly separable from dependence
assigned to the other. Because \textsc{Delphes} fixes $\mu$ per sample, a mock
value $\mu'$ is constructed with a correlation to $N_{\rm PV}$ that can be set
directly,
\begin{equation}
  \mu' \;=\; \rho_{\mu N}\,\mu_{\mathrm{phys}}(N_{\mathrm{PV}})
  \;+\; (1 - \rho_{\mu N})\,\mu_{\mathrm{indep}},
\end{equation}
where the correlated piece $\mu_{\mathrm{phys}}$ is drawn from a normal
distribution of mean $2N_{\rm PV}$ and width $\sqrt{2N_{\rm PV}}$, the
uncorrelated piece $\mu_{\mathrm{indep}}$ is drawn uniformly from $[1,120]$ and
the mixing factor $\rho_{\mu N}$ sets the degree of correlation.~\footnote{Note that strictly speaking $\mu$ relates to the experimental luminosity through the inelastic cross-section, while $N_{\rm PV}$ is also sensitive to the vertex reconstruction efficiency. Without loss of generality, we implement here an effective correlation among these quantities.}
Seven values,
$\rho_{\mu N} \in \{0, 0.1, 0.3, 0.5, 0.7, 0.8, 1.0\}$, are used, spanning
independent to fully correlated; examples are shown in
Fig.~\ref{fig:pu_corr_sim}.

The dependence itself is added to the reconstructed transverse momentum of each
jet as
\begin{equation}
    p_{\textrm{T}}^{\mathrm{reco}} \;\to\; p_{\textrm{T}}^{\mathrm{reco}}
    + \alpha(p_{\textrm{T}}^{\mathrm{gen}}, |\eta|)\,(N_{\mathrm{PV}} - 1)
    + \beta (p_{\textrm{T}}^{\mathrm{gen}}, |\eta|)\,\mu',
\end{equation}
with the coefficients carrying the dependence on the jet through
\begin{align}
    \alpha(p_{\textrm{T}}, |\eta|) &= (a_0 + a_1 \ln p_{\textrm{T}})\,(1 + a_2 |\eta|), \\
    \beta (p_{\textrm{T}}, |\eta|) &= (b_0 + b_1 \ln p_{\textrm{T}})\,(1 + b_2 |\eta|).
\end{align}
Both coefficients are evaluated at the particle-level transverse momentum, so
that the injected dependence is a fixed function of the true jet rather than of
the quantity being corrected. The first term is proportional to
$N_{\mathrm{PV}} - 1$ and so vanishes for an event with a single vertex, in which
there is no pile-up to remove.

Two sets of coefficients are used, named lo-$p_{\textrm{T}}$ and hi-$p_{\textrm{T}}$,  given in Tab.~\ref{tab:pu_coeffs}. They differ
only in the strength of the logarithmic transverse-momentum dependence, which is
five times larger in the hi-$p_{\textrm{T}}$ regime than in the lo-$p_{\textrm{T}}$ regime. In both,
the injected dependence is linear in $N_{\rm PV}$ and in $\mu'$, which is the
functional form the parametric correction of \secref{sec:pileup_atlas} assumes. An example of the injected dependence is given in Fig~\ref{fig:pu_inject}. More complex examples which do not match the functional form of the parametric approach are explored in subsequent subsections.

\begin{table}[]
    \centering
    \caption{Parameters for the $\alpha$ and $\beta$ functions which control the \pt\ and $\eta$ dependence of the injected pileup.}
    \begin{tabular}{c |ccc|ccc}
    \hline\hline
        regime & $a_0$ & $a_1$ & $a_2$ & $b_0$ & $b_1$ & $b_2$    \\
            \hline
        lo-$p_{\textrm{T}}$ & \multirow{2}{*}{0.4} & 0.08 & \multirow{2}{*}{0.15} & \multirow{2}{*}{0.15} &0.04 & \multirow{2}{*}{0.10} \\
        hi-$p_{\textrm{T}}$ &   & 0.40 & & & 0.20\\
            \hline\hline
    \end{tabular}
    \label{tab:pu_coeffs}
\end{table}

A distortion of the absolute energy scale is applied to these samples as well. It
is injected before the pile-up correction, so that this stage operates on jets
whose absolute energy scale is already miscalibrated, but it is described in \secref{sec:abs}
where it is corrected.

\subsection{\texorpdfstring{Parametric method}{Parametric method}}
\label{sec:pileup_atlas}

The baseline correction follows the procedure used by
ATLAS~\cite{ATLAS:2020cli}, which has two components, a subtraction based on the
jet area and a residual correction derived from simulation.

The first component removes the pile-up energy expected in the region the jet
occupies. The pile-up density $\rho$ is the median of $p_{\textrm{T}}/A$ over the jets of
the event, computed in ATLAS from $k_t$ jets of radius $0.4$ built from
positive-energy calorimeter clusters within $|\eta| < 2$~\cite{ATLAS:2015ull}. The $k_t$ algorithm tends to collect uniform soft
background into its jets, and the median is used in place of the mean because it
is insensitive to the few hard jets which populate the tail of the $p_{\textrm{T}}/A$
distribution. The product $\rho \times A$ is removed as a scale factor on the jet
four-momentum, so the jet direction is unchanged.

The measured $\rho$ falls
to nearly zero beyond $|\eta| \simeq 2$, because the coarser segmentation of the
forward calorimeter lowers the occupancy there, so a median taken over the full
pseudorapidity range would describe the central region less well than one taken
over the central region alone. The single central value is then applied to jets
at all pseudorapidities,  over‑correcting central jets and under‑correcting forward ones. DELPHES does not fully reproduce this effect; a stronger version is modeled directly in \secref{sec:etadens}.

A central estimate of $\rho$ describes neither the pile-up sensitivity of the
forward calorimeter nor that of the dense core of a high-$p_{\textrm{T}}$ jet, so some
dependence on the pile-up conditions survives the subtraction. A residual
correction is therefore applied, giving
\begin{equation}
  p_{\textrm{T}}^{\mathrm{corr}} = p_{\textrm{T}}^{\mathrm{reco}} - \rho \times A
  - \alpha \times (N_{\mathrm{PV}} - 1) - \beta \times \mu ,
  \label{eq:atlas_pu}
\end{equation}
where $\alpha$ carries the dependence on the number of reconstructed primary
vertices and $\beta$ that on the mean number of interactions per bunch crossing.

The two coefficients $\alpha$ and $\beta$ are derived by ATLAS in bins of $p_{\textrm{T}}^{\mathrm{true}}$ and
$|\eta_{\mathrm{det}}|$. Within a bin, the two coefficients cannot be trivially obtained by regressing on each
variable separately, because $N_{\mathrm{PV}}$ and $\mu$ are correlated in the
sample. Such a fit returns not $\alpha$ but $\alpha$ plus $\beta$ times the slope of
$\mu$ against $N_{\mathrm{PV}}$, so the $\mu$ dependence leaks into $\alpha$
through the shared variation, and applying both corrections would count it twice. ATLAS removes the leak by conditioning. Within each
$(p_{\textrm{T}}^{\mathrm{true}}, |\eta_{\mathrm{det}}|)$ bin the jets are divided further
in $\mu$, and the $N_{\mathrm{PV}}$ slope is measured separately in each. Because $\mu$ varies little across a division, the slope measures $\alpha$ alone. The $\mu$
coefficient is obtained the same way, dividing instead in $N_{\mathrm{PV}}$.
The slopes
are averaged to give the
coefficient for that bin.  Averaging is valid  if $\alpha$ is the same in every division, that is
if $\partial r / \partial N_{\mathrm{PV}}$ does not itself depend on $\mu$, which is assumed above and in the ATLAS approach.  The assumption is examined  in
\secref{sec:pu_cross}, where an interaction term is injected.

Both coefficients vary logarithmically with $p_{\textrm{T}}^{\mathrm{true}}$, so a
logarithmic fit is performed over $20 < p_{\textrm{T}}^{\mathrm{true}} < 200$~GeV in each
bin of $|\eta_{\mathrm{det}}|$ and evaluated at $p_{\textrm{T}}^{\mathrm{true}} = 25$~GeV,
the region in which pile-up matters most. The resulting values are fit linearly
against $|\eta_{\mathrm{det}}|$, which suppresses statistical fluctuations and
allows the coefficients to be sampled smoothly across the detector.


\subsection{NN Method}

\subsubsection{Network structure}

The correction is produced by a multilayer perceptron, a feed-forward network of
fully connected layers, which takes four quantities per jet. Two
describe the jet, the logarithm of the transverse momentum remaining after the
area-based subtraction and the signed $\eta_{\mathrm{det}}$, and two describe the
event, the number of reconstructed primary vertices $N_{\mathrm{PV}}$ and the
mean number of interactions per bunch crossing $\mu$. The network returns a
single number $\delta$, in GeV, which is subtracted from the area-subtracted
transverse momentum. The pseudorapidity is supplied signed rather than folded
about zero, so that the network is free to describe the two halves of the
detector differently, where the parametric coefficients of \secref{sec:pileup_atlas}
are functions of $|\eta_{\mathrm{det}}|$ alone.

Each input is standardized using the mean and width of the training set. The
network has four hidden layers of width 256 with exponential linear unit
activations. The final layer is initialized near zero, so that training begins
from a network which applies no correction at all.

\subsubsection{Covariance loss}

This stage is asked to remove the dependence of the residual on the pile-up
conditions, not to set its absolute scale, which the correction of
\secref{sec:abs} sets. A network trained against the particle-level
transverse momentum of each jet would target both at once. The network is
therefore trained directly against the quantity this stage is meant to remove,
through a loss built from the covariance of the corrected residual
\begin{equation}
    r \;=\; p_{\textrm{T}}^{\mathrm{reco}} - p_{\textrm{T}}^{\mathrm{gen}} - \rho A - \delta
\end{equation}
with the variables which describe the pile-up. A correction which leaves $r$
uncorrelated with those variables satisfies this stage regardless of what value
$r$ itself takes.

The two pile-up variables are standardized, and $\mu$ is then orthogonalized
against $N_{\mathrm{PV}}$,
\begin{equation}
    \mu_\perp \;=\; \mu_s \;-\;
    \frac{\mathrm{Cov}(\mu_s,\, N_{\mathrm{PV},s})}{\mathrm{Var}(N_{\mathrm{PV},s})}\,
    N_{\mathrm{PV},s},
\end{equation}
so that the term built from $\mu$ measures only the dependence which
$N_{\mathrm{PV}}$ does not already describe and the shared dependence is not
counted twice. This matters most in the samples with large $\rho_{\mu N}$, where
the two variables carry nearly the same information.

The loss is a sum of five covariance terms and one penalty. The first two
penalize linear dependence of the residual on $N_{\mathrm{PV}}$ and on
$\mu_\perp$. The next two use the squares of those variables and so penalize
curvature in the residual profile, which a linear term cannot see. The fifth uses
their product and penalizes dependence on the two jointly. The penalty acts on
negative values of $\delta$, since in-time pile-up adds energy to a jet and a correction
which removes energy is therefore expected to be positive, and is applied through
the softplus function so that it remains smooth,
\begin{equation}
\begin{split}
    \mathcal{L} \;=\;
        &\,\mathrm{Cov}(r, N_{\mathrm{PV},s})^2
        + \mathrm{Cov}(r, \mu_\perp)^2 \\
        &+ \mathrm{Cov}(r, N_{\mathrm{PV},s}^{\,2})^2
        + \mathrm{Cov}(r, \mu_\perp^{\,2})^2 \\
        &+ \mathrm{Cov}(r, N_{\mathrm{PV},s}\,\mu_\perp)^2
        + \lambda\,\overline{\mathrm{softplus}(-\delta)}.
\end{split}
\end{equation}

A loss built only from these terms can be made small by cancellation, leaving a
positive slope in one region of the detector against a negative slope in another.
The same linear covariance terms are therefore evaluated separately in each bin
of $|\eta|$ and of $p_{\textrm{T}}$ and added to the loss,
\begin{equation}
    \mathcal{L} \;\mathrel{+}=\;
        \sum_b w_b \left[\,
            \mathrm{Cov}(r_b, N_{\mathrm{PV},b})^2
            + \mathrm{Cov}(r_b, \mu_{\perp,b})^2
        \,\right],
        \qquad
        w_b = \frac{\sqrt{N_b}}{\langle {\sqrt{N_b} \rangle}},
\end{equation}
each covariance divided by the spread of the pile-up variable within the bin. The weight $w_b$ grows as the
square root of the number of jets in the bin, so the densely populated central
bins carry the most weight while sparse forward and high-$p_{\textrm{T}}$ bins still
contribute. Here $N_b$ is the number of jets of the minibatch in bin $b$, the average is over the bins.

\subsubsection{Training strategy}

Events with little pile-up outnumber the rest, and it is the events with the most
pile-up which carry the dependence the loss is built to measure. Minibatches are
therefore drawn from a $10\times10$ grid in $(N_{\mathrm{PV}}, \mu)$ with each
cell sampled equally, so that the sparsely populated high-pile-up cells
contribute nearly as much to the covariance as the populated bulk.

Training uses Adam~\cite{kingma2017adammethodstochasticoptimization} with a learning rate of $3 \times 10^{-4}$, a batch
size of 4096 and a penalty weight $\lambda = 0.1$, for a budget of 150 epochs,
with gradients clipped at a norm of 5. The learning rate follows a cosine
schedule with warm restarts~\cite{SGDR}, falling to one hundredth of its initial value over a
cycle of one third of the budget before returning to it. Training is stopped
early on $|\mathrm{Cov}(r, N_{\mathrm{PV}})| + |\mathrm{Cov}(r, \mu)|$ measured
on held-out jets, the same quantity the loss is built from rather than the loss
itself, once that quantity has not improved for 50 epochs. The criterion is
applied only after the first third of the budget, so that a restart in the
learning-rate schedule is not mistaken for a failure to improve.

\subsection{Performance}
\label{sec:pileup_perf}

\begin{figure}[ht!]
    \centering
    \includegraphics[width=0.85\linewidth]{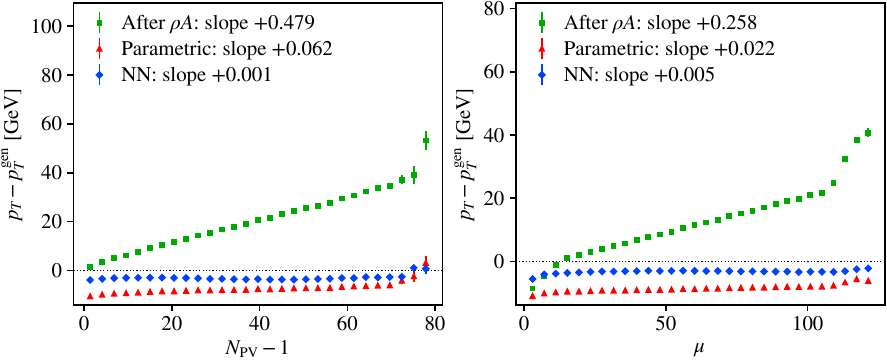}
    \includegraphics[width=0.85\linewidth]{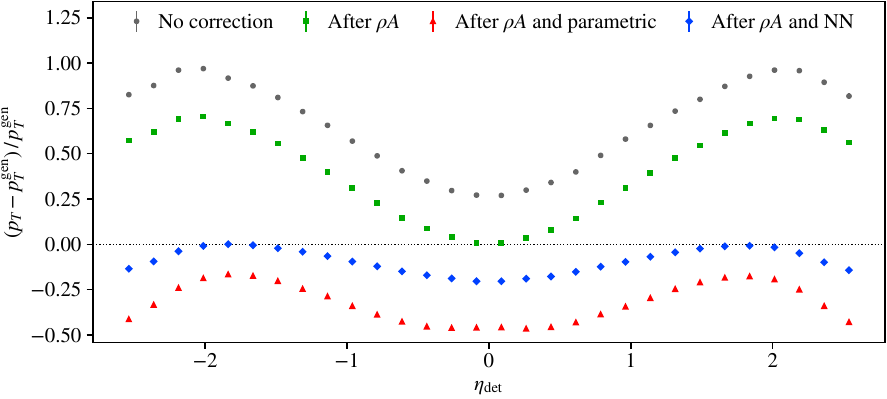}
     \includegraphics[width=0.85\linewidth]{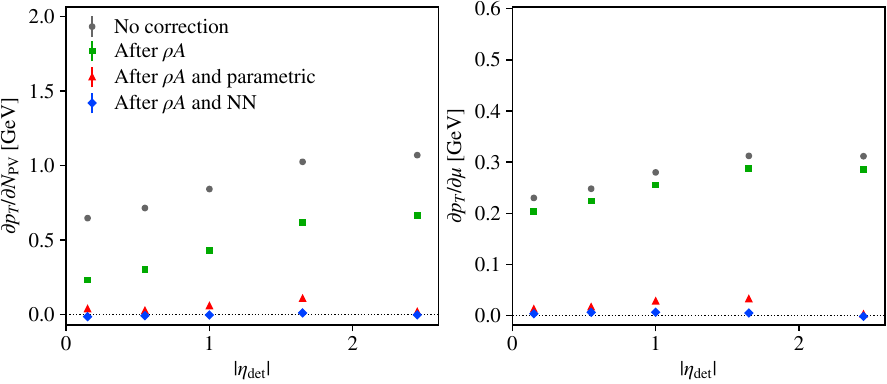}
    \caption{Performance of the NN-based and parametric pileup corrections for one example dataset, in the lo-$p_{\textrm{T}}$ regime with $\rho_{\mu N}=0.1$}
    \label{fig:pu_corr}
\end{figure}

The metric for the correction is the remaining dependence of the residual on $ N_{\mathrm{PV}}$ and $\mu$, 
the slopes $\partial r / \partial N_{\mathrm{PV}}$ and
$\partial r / \partial \mu$. Figure~\ref{fig:pu_corr}a shows these slopes before
and after correction for one dataset. 
Figure~\ref{fig:pu_corr}b shows the relative residual against $\eta_{\mathrm{det}}$ before correction, after the area correction and after each of the two corrections. The residual itself is non-zero, as only the slopes are targeted by this stage. The slopes are also measured in  bins of $|\eta_{\mathrm{det}}|$, as shown in
Figure~\ref{fig:pu_corr}c, and the overall figure of merit is the root mean
square of these per-bin slopes, in GeV per unit $N_{\mathrm{PV}}$ or per unit
$\mu$.

\begin{figure}[ht!]
    \centering
    \includegraphics[width=0.45\linewidth]{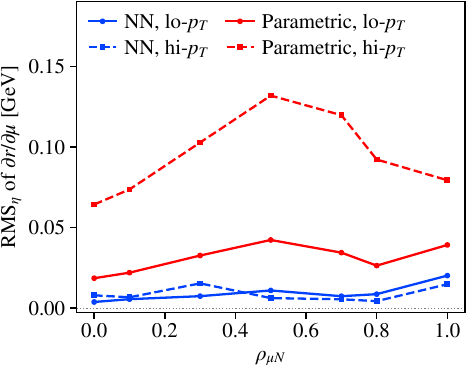}
        \includegraphics[width=0.45\linewidth]{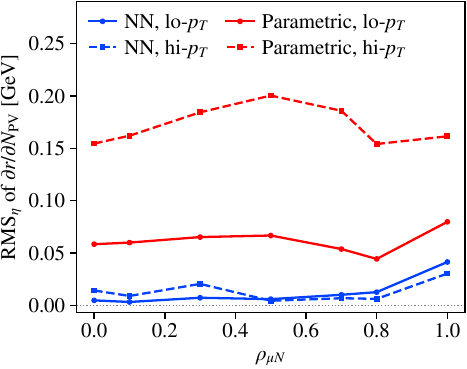}
    \caption{Performance metrics for PU dependence removal versus injected correlation $\rho_{\mu N}$}
    \label{fig:pu_corr_scan}
\end{figure}

Figure~\ref{fig:pu_corr_scan} shows both figures of merit against the injected
correlation $\rho_{\mu N}$, for each method and each $p_{\textrm{T}}$ regime. The residual
dependence left by the parametric baseline varies with the correlation between the two pile-up variables, while the NN method's performance is
comparatively flat across the scan. Table~\ref{tab:pu} averages each figure of merit over the seven values of
$\rho_{\mu N}$. 

The advantage is larger in the hi-$p_{\textrm{T}}$ regime, as expected, due to the parametric baseline approach collapsing the transverse-momentum dependence of each
coefficient to the value at $p_{\textrm{T}}^{\mathrm{true}} = 25$~GeV and the dependence absorbed
into a systematic uncertainty, while the network takes the
jet transverse momentum as an input and is under no such constraint. In this test, the form of the injected dependence respects the assumption of 
linearity in $N_{\mathrm{PV}}$ and in $\mu$ made by the parametric baseline; see Eq.~(\ref{eq:atlas_pu}). The larger gains, here in the hi-$p_{\textrm{T}}$ regime and in the appendix for a forward pile-up density, come from dependence outside the functional form of the baseline, which is the point being demonstrated, and are not a claim that the baseline is poorly tuned.

The covariance loss constrains only the dependence of the residual on the pile-up variables, not its additive offset, so after the NN correction the residual sits away from zero (Figs.~\ref{fig:etadens_perf}--\ref{fig:pu_cross_resid}); the offset is absorbed by the absolute correction of \secref{sec:abs}. A production version could add a term anchoring the offset.

In Appendices~\ref{sec:etadens}--\ref{sec:pu_cross}, we study the performance for three further injections: a mis-estimated forward pile-up density, a non-linear dependence on $N_{\mathrm{PV}}$ and an interaction between the two pile-up variables.        

\begin{table}[ht]
\centering
\caption{Residual pile-up dependence after correction, for the lo- and hi-pT regimes, for NN and parametric methods.  For each dataset, the RMS over $\eta$ bins of the residual slopes ($\partial r / \partial N_{\mathrm{PV}}$ and $\partial r / \partial \mu$) is measured. Shown is the mean of this RMS across the several choices of $\rho_{\mu N}$. The NN/Param. ratio is the residual dependence the NN
leaves relative to the baseline.}
\label{tab:pu}
\begin{tabular}{l ccc ccc}
\hline
 & \multicolumn{3}{c}{$\partial r / \partial N_{\mathrm{PV}}$}
 & \multicolumn{3}{c}{$\partial r / \partial \mu$} \\
\cmidrule(lr){2-4} \cmidrule(lr){5-7}
$p_T$ regime & NN & Param. & NN/Param. & NN & Param. & NN/Param. \\
\hline
lo-$p_T$ & 0.012 & 0.061 & 0.20 & 0.009 & 0.031 & 0.30 \\
hi-$p_T$ & 0.013 & 0.172 & 0.08 & 0.009 & 0.095 & 0.09 \\
\hline
all      & 0.013 & 0.116 & 0.11 & 0.009 & 0.063 & 0.14 \\
\hline
\end{tabular}

\end{table}

\subsection{Summary of Pile-up Corrections}

The NN-based method is more flexible and general than the parametric method, able to accommodate dependence in transverse momentum, non-linear dependence and interaction between pile-up terms.

\section{Absolute energy scale}
\label{sec:abs}

The second stage of the calibration corrects the average response of the
detector, so that jet energies on average agree with the energies of their particle-level jets. This section describes the distortion injected to make that
correction a realistic task, the parametric method against which the network is
compared, the network itself and the metrics on which the two are judged.

\subsection{Injection of absolute response}

The response of the simulated detector, the ratio of the reconstructed to the
particle-level jet energy, is shown in Fig.~\ref{fig:jet_res}. It lies near or above unity,
since pile-up has not yet been subtracted there. A real calorimeter behaves differently: the response of an
ATLAS jet before calibration is well below unity, rises with energy and falls in
the forward region~\cite{ATLAS:2020cli}. To make the task more realistic, a multiplicative distortion of the jet energy is applied,
\begin{equation}
  \mathcal{R}(E,|\eta|) = r_0\,
    \bigl(1 - r_1\,e^{-E/E_0}\bigr)\,
    \bigl(1 - r_2\,[\max(|\eta|-\eta_0,\,0)]^2\bigr),
  \label{eq:abs_inject}
\end{equation}
evaluated at the particle-level energy
$E^{\mathrm{gen}} = p_{\textrm{T}}^{\mathrm{gen}}\cosh\eta^{\mathrm{gen}}$, so that it is a
fixed function of the true jet. It
multiplies the reconstructed energy left by the pile-up injection of
\secref{sec:pileup}, and the transverse momentum is recomputed from the distorted
energy with the jet mass unchanged.

The three factors of Eq.~(\ref{eq:abs_inject}) reproduce three features of a real
calorimeter. The overall scale $r_0 = 0.90$ is a constant deficit of $10\%$,
standing for the lower signal a non-compensating calorimeter registers from a
hadronic shower than from an electromagnetic one of the same energy. The
exponential factor, of depth $r_1 = 0.65$ and characteristic energy
$E_0 = 35$~GeV, suppresses the response further at low energy, to describe losses in
material in front of the calorimeter, and energy falling outside the jet cone. The last
factor is flat within $|\eta| < \eta_0 = 0.7$ and falls quadratically beyond it
with strength $r_2 = 0.08$, degrading the response in the forward region.

\subsection{\texorpdfstring{Parametric method}{Parametric method}}
\label{sec:abs_atlas}

ATLAS brings the jet to the particle-level scale with two corrections, one acting
on the energy and one on the reconstructed direction~\cite{ATLAS:2020cli}. The
first accounts for the non-compensating response of the calorimeter, for energy
lost in the material in front of it and for energy falling outside the jet cone.
The second removes a bias in the reconstructed pseudorapidity, which appears when
a jet straddles two calorimeter regions of different response, so that one side
of the jet is weighted more heavily than the other. Both are derived from
simulated dijet events to which the pile-up corrections have already been
applied.

The average response $\mathcal{R}$ is measured in bins of $E^{\mathrm{true}}$ and
$\eta_{\mathrm{det}}$, as the mean of a Gaussian fit to the core of the
$E^{\mathrm{reco}}/E^{\mathrm{true}}$ distribution.  The correction must act on the measured energy, while the response has been
measured against the true energy, which is not available when the correction is
applied, requiring numerical inversion~\cite{Cukierman:2016ntt,ATLAS:2018gni}. The response is evaluated
on a grid of true energies, each point is carried to the reconstructed energy it
implies through $E^{\mathrm{reco}} = E^{\mathrm{true}}\,\mathcal{R}$, and the
calibration factor $1/\mathcal{R}$ is tabulated against $E^{\mathrm{reco}}$.

The implementation used here follows the same sequence. Jets are divided into
nine slices of $|\eta_{\mathrm{det}}|$ placed at the transitions between
calorimeter regions, and binned logarithmically in $E^{\mathrm{true}}$ within
each. The response in each bin is the mean of an iterative Gaussian fit to the
core of the distribution. The profiled points are smoothed by a cubic spline in
$\log E^{\mathrm{true}}$, weighted by the uncertainty on each point, and
re-interpolated monotonically so that $E^{\mathrm{true}}\,\mathcal{R}$ increases
everywhere and the mapping to $E^{\mathrm{reco}}$ can be inverted.

Relative to the published procedure, there are two differences in our implementation. The correction for the
bias in reconstructed pseudorapidity is absent, and the smoothing spline uses an
automatically chosen smoothing parameter, so its effective length varies with the
number of points in the fit.

\subsection{NN method}
\label{sec:abs_nn}

\subsubsection{Response model}

The network learns the response
$\mathcal{R} = E^{\mathrm{reco}}/E^{\mathrm{true}}$ as a function of the true
energy and the pseudorapidity, but as a continuous function of both rather than
as a value per bin, following the broader idea of parametrizing detector response with neural networks~\cite{DetectorResponseNN}. The median is targeted directly, using a quantile loss at
$\tau = 0.5$, which is the mean absolute deviation
\begin{equation}
  \mathcal{L} \;=\; \frac{1}{2N}\sum_{i=1}^{N}
    \bigl| \mathcal{R}_i - \mathcal{R}^{\mathrm{pred}}_i \bigr| .
\end{equation}

The network is given six inputs:  $\log E^{\mathrm{true}}$ and
$|\eta_{\mathrm{det}}|$, as well as $1/E$ and $1/E^2$ for the behaviour at low energy, $|\eta|^2$ for the degradation in the forward
region and $\log E \cdot |\eta|$ for interdependence. These four are functions of the first two and are supplied to aid convergence.

The architecture is nine hidden layers of width 384 with sigmoid linear unit
activations, residual skip connections from the second layer onward and no dropout. The single output is passed through a softplus function
normalized so that a zero input gives unity, and the final layer is initialized at
zero, so the response begins at $\mathcal{R} = 1$ everywhere and remains positive
throughout training.

Jets are drawn for training with weights which partly flatten the distribution in
$(\log E^{\mathrm{true}}, |\eta_{\mathrm{det}}|)$ on a $25 \times 12$ grid, using
$w \propto 1/\sqrt{N}$ in each cell.
Optimization uses AdamW~\cite{AdamW} with a weight decay of $5 \times 10^{-3}$ and a one-cycle learning
rate peaking at $3 \times 10^{-3}$, for 100 epochs with a batch size of 4096,
gradients clipped at a norm of 5 and $15\%$ of the jets held out for validation.

\subsubsection{Numerical inversion}

The network learns the response against the true energy, and the correction must be applied to the measured
energy. The inversion of \secref{sec:abs_atlas} is therefore used again and is
the same in both methods, so that the comparison between them rests on the
response model alone. Within each slice of $|\eta_{\mathrm{det}}|$, the response
is evaluated on a grid of  true energies, each point is carried to the
reconstructed energy it implies, and the calibration factor
$1/\mathcal{R}$ is tabulated against it and interpolated at inference.

 The response returned by the network can carry
small local wiggles, left by the stochastic nature of the training, which are removed with a Savitzky--Golay filter, a
local polynomial smoother, applied to the grid before inversion, where the
baseline uses a smoothing spline for the same purpose at an earlier stage. The
reconstructed energies are then required to increase along the grid, since a
lookup table can only be built from a monotonic mapping. Because the response
learned here varies continuously with pseudorapidity, the tables are rebuilt on a
grid finer than the slices used to bin the baseline, in steps of $0.05$ in
$|\eta_{\mathrm{det}}|$.

\subsection{Performance}
\label{sec:abs_perf}

\begin{figure}
    \centering
    \includegraphics[width=0.85\linewidth]{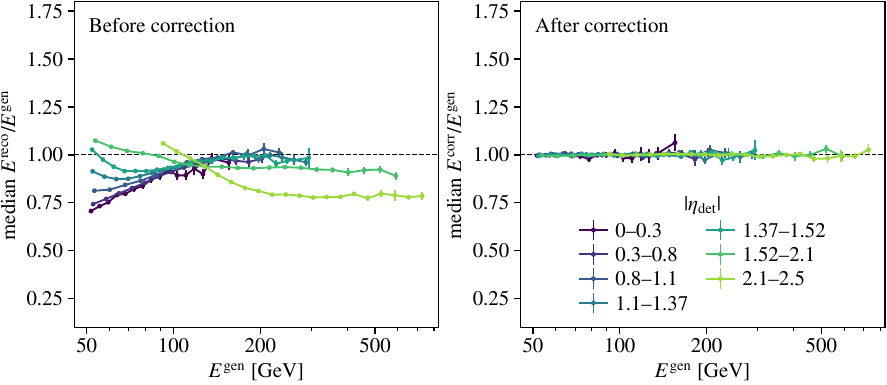}
        \includegraphics[width=0.85\linewidth]{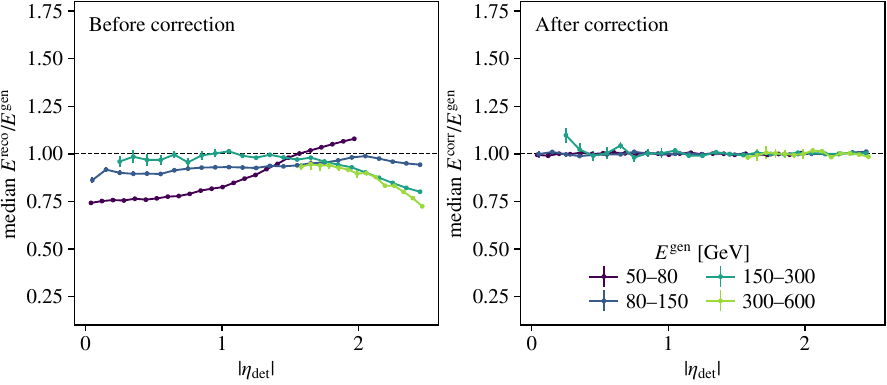}

    \caption{ Performance of the NN-based absolute energy scale correction. Shown is the median response before correction, $E^{\textrm{reco}}/E^{\textrm{gen}}$ (left), and after, $E^{\textrm{corr}}/E^{\textrm{gen}}$ (right), versus $E^{\textrm{gen}}$ (top) or detector pseudorapidity $\eta$ (bottom).}
    \label{fig:abs}
\end{figure}

The performance of both methods is measured by assessing the deviation of the median of the corrected response,
$E^{\mathrm{corr}}/E^{\mathrm{gen}}$, from unity, in bins of $E^{\mathrm{gen}}$ and
$|\eta_{\mathrm{det}}|$, for jets with $E^{\mathrm{gen}}>50$ GeV, to stay clear of the reconstruction threshold bias.   Figure~\ref{fig:abs} shows the performance for one dataset, the response against
energy in slices of pseudorapidity and against pseudorapidity in slices of
energy, before and after correction. Note that the response shape before correction is partially due to pileup; the pileup correction step removes the dependence on $\mu$ and $N_{\textrm{PV}}$ but leaves the offset for this stage to handle.

\begin{figure}
    \centering
    \includegraphics[width=0.95\linewidth]{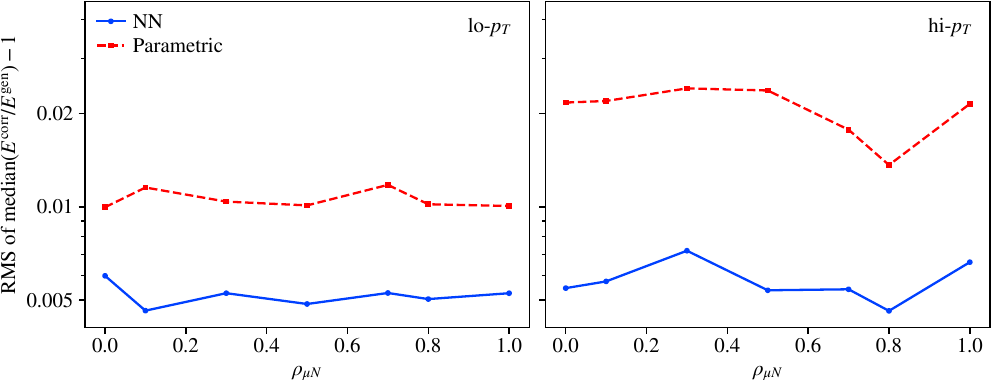}
    \caption{ RMS of the deviation of the median response from unity  over a 2D grid in energy and pseudorapidity, for the lo-$p_\textrm{T}$ (left) and hi-$p_\textrm{T}$ datasets (right), as a function of the correlation $\rho_{\mu N}$.}
    \label{fig:abs_closure}
\end{figure}

For a quantitative summary, we calculate the   root-mean-squared across bins, which weights the largest deviations most and prevents bins
of opposite sign from canceling.  This summary statistic is calculated in three ways: over bins of $E^{\mathrm{gen}}$,  bins of $|\eta_{\mathrm{det}}|$ and over a 2D grid of both.   Figure.~\ref{fig:abs_closure} shows the summary statistic for each of the datasets, in the lo-$p_\textrm{T}$ and hi-$p_\textrm{T}$ regimes, across values of $\rho_{\mu N}$, compared to the parametric baseline. Both corrections are derived and applied on the output of the NN pile-up
correction of \secref{sec:pileup}. Table~\ref{tab:abs_closure} shows the performance averaged over the seven values of $\rho_{\mu N}$
 in each regime. 

\begin{figure}
    \centering
     \includegraphics[width=0.95\linewidth]{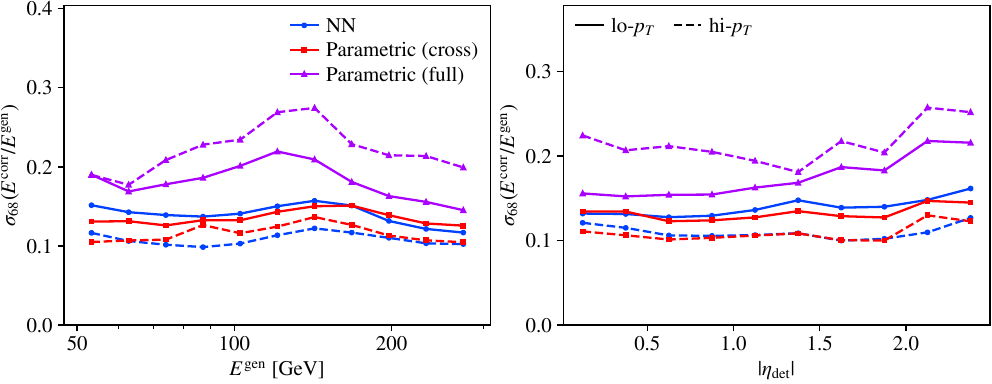}
    \caption{Comparison of parametric and NN performance for absolute corrections in terms of resolution versus energy (left) or pseudorapidity (right), for $\rho_{\mu N}=0.1$. The parametric approach is shown for two scenarios: one which uses the output of the NN pileup correction ({\it cross}), and another that uses the output of the parametric pileup correction ({\it full}).}
    \label{fig:abs_sigma68}
\end{figure}

The resolution is reported as well in Fig.~\ref{fig:abs_sigma68}, as half
the central interval of the corrected response,
\begin{equation}
  \sigma_{68} \;=\; \tfrac{1}{2}\Bigl[\,Q_{0.8413}(\mathcal{R})
  \;-\; Q_{0.1587}(\mathcal{R})\,\Bigr],
\end{equation}
where $Q_p$ is the $p$-th quantile and the two quantiles are those a Gaussian
places at $\pm 1$ standard deviation.

Note that the resolution itself is not the target of this stage, where a single factor is derived from the jet energy and pseudorapidity, targeting the median rather than the spread of the
response at fixed energy and pseudorapidity. Narrowing that spread requires
features which vary from jet to jet with the resolution itself, which is the
purpose of the global sequential calibration as well as the sub-jet calibration described below. At the lowest energies, the parametric correction gives a slightly narrower resolution than the network (Fig.~\ref{fig:abs_sigma68}), while the network closes the median better (Table~\ref{tab:abs_closure}). At fixed true energy, the width of the corrected response depends on how steeply the correction varies with the measured energy.

\begin{table}[ht]
\centering
\caption{Performance of the absolute energy-scale correction, NN versus the parametric baseline, averaged over the seven $\rho_{\mu N}$ values in each $p_{\textrm{T}}$ regime. Entries are the RMS across bins of $\mathrm{median}(E^{\mathrm{corr}}/E^{\mathrm{gen}}) - 1$; zero is ideal.  The RMS is calculated in three ways: over bins of $E^{\mathrm{gen}}$,  bins of $|\eta_{\mathrm{det}}|$ and over a 2D grid of both.}
\label{tab:abs_closure}
\begin{tabular}{l ccc ccc}
\hline \hline
 & \multicolumn{3}{c}{NN} & \multicolumn{3}{c}{Param.} \\
\cmidrule(lr){2-4} \cmidrule(lr){5-7}
$p_{\textrm{T}}$ regime & $E_{\rm gen}$ & $|\eta|$ & 2D & $E_{\rm gen}$ & $|\eta|$ & 2D \\
\hline
lo-$p_T$ & 0.0022 & 0.0011 & 0.0052 & 0.0100 & 0.0070 & 0.0106 \\
hi-$p_T$ & 0.0037 & 0.0025 & 0.0058 & 0.0172 & 0.0020 & 0.0206 \\
\hline \hline
\end{tabular}
\end{table}

\section{Sub-jet--resolved energy scale}
\label{sec:subjet_y}

\begin{figure}
    \centering
    \includegraphics[width=0.35\linewidth]{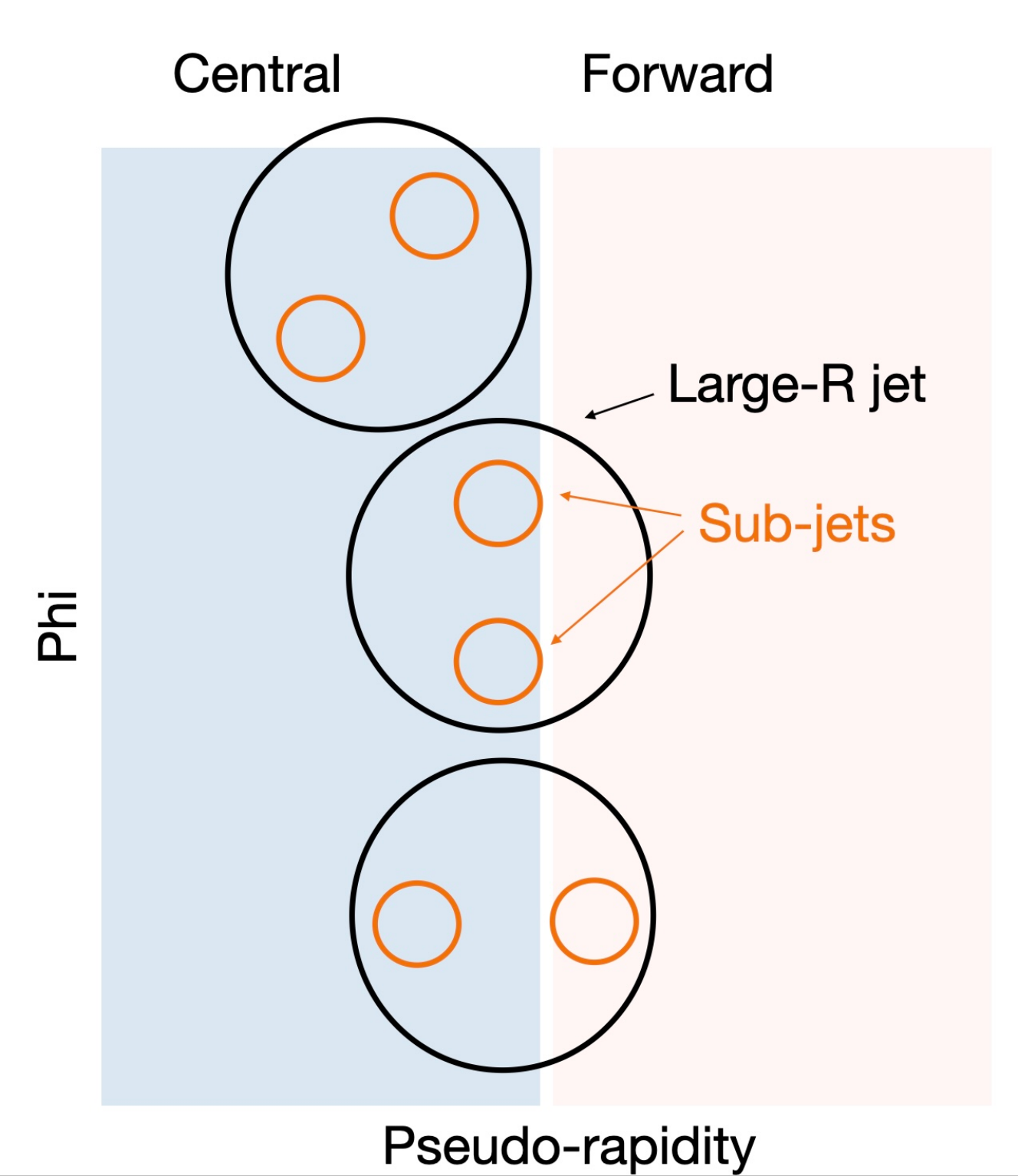}
    \caption{For large-radius jets with most of their energy in subjets, the response at the same $|\eta|$ can depend on the subjet axis angle where the response varies with pseudorapidity, as it does near cracks.}
    \label{fig:subjet}
\end{figure}

Neural-network calibrations of large-radius jet energy and mass have been demonstrated by ATLAS~\cite{ATLAS:2023dnn}. Here, the NN method allows for inclusion of additional jet features without designing new parametric forms. We demonstrate this flexibility by adding internal jet sub-structure information to improve the calibration of jets which straddle a central-forward crack. Two jets at the same pseudo-rapidity and energy can require quite different corrections depending on where their subjets fall relative to the crack. Figure~\ref{fig:subjet} illustrates the issue. The quantity which distinguishes the two cases is the orientation of the axis
joining the sub-jets, written ${\sjangle} = \operatorname{atan2}(\Delta\phi, \Delta\eta)$,
so that ${\sjangle} \in (-\pi, \pi]$
is the angle of the axis measured from the $\eta$ direction towards the $\phi$ direction.

The jet longitudinal angle is described via rapidity $|y|$ when describing the large-$R$ jet response and performance~\cite{Gallicchio:2018elx}. The detector-relative quantities, the position of the transition, the positions of the sub-jets and the conversions between transverse momentum and energy, are described via detector $\eta$.

The sample is described above (\secref{sec:samples}), photon plus hadronically
decaying $W$ events, clustered  with radius $R = 1.0$. Each large-$R$ jet carries
its two leading soft-drop sub-jets, at reconstructed and at particle level, and
those sub-jets drive both the distortion injected below and the correction
derived against it.

The distortion of \secref{sec:abs} is applied with a step added at a fixed
position in the detector,
\begin{equation}
  \mathcal{R}(E,|\eta|) = r_0\bigl(1-r_1 e^{-E/E_0}\bigr)
  \bigl(1-c_d\,\sigma\!\bigl((|\eta|-\eta_c)/c_w\bigr)\bigr),
\end{equation}
with $\sigma$ the logistic function. The response falls by $c_d$ beyond
$|\eta| = \eta_c$ over a width $c_w$, and the quadratic forward roll-off of
\secref{sec:abs} is switched off, so that the step is the only structure in
pseudorapidity and its effect is not confused with a general degradation of the
forward response. The parameters are $\eta_c = 1.37$, a depth $c_d = 0.45$ and a
width $c_w = 0.05$, on an overall scale $r_0 = 0.90$ with a low-energy turn-on of
depth $r_1 = 0.30$ and characteristic energy $E_0 = 300$~GeV, the turn-on being
placed at a higher energy than for the small-radius jets because these jets are
more energetic. Pile-up is injected as in \secref{sec:pileup} with the two
pile-up variables uncorrelated.

The distortion is applied to the components of the jet, depending on their specific detector locations.  The jet is divided
into three, the two sub-jets and the energy which belongs to neither,
$E_{\mathrm{jet}} - E_{\mathrm{sjA}} - E_{\mathrm{sjB}}$. Each sub-jet is
distorted at its own energy and its own position, $(E_{\mathrm{sj}},
|\eta_{\mathrm{sj}}|)$; the remainder is distorted at the energy and the axis position of the large-$R$ jet,
$\mathcal{R}(E_{\mathrm{jet}}, |\eta_{\mathrm{jet}}|)$. The three distorted energies are summed to give the distorted jet energy, while the
direction and the mass of the jet are kept at their reconstructed values: the distortion
changes the energy of the jet and not its axis.  The pile-up shift is divided between the
sub-jets in proportion to their share of the transverse momentum, a proxy for the share of the jet area that the Delphes output does not
include for sub-jets; 
the sub-jet momenta are not inputs to the correction.

The pile-up correction of \secref{sec:pileup} is applied first, so that both
corrections here act on jets from which the pile-up dependence has already been
removed.



The network is the response model of \secref{sec:abs_nn} but with additional inputs. Four inputs describe the jet alone: $\log E^{\mathrm{true}}$, the large-$R$-jet
coordinate $|y|$, its square and the product of the two.  Three inputs describe the sub-jets: two
are the positions of the sub-jets, $|\eta_{\mathrm{sj,hi}}|$ and
$|\eta_{\mathrm{sj,lo}}|$, ordered so that the first is the more forward of the
pair. The third is the
distance from the step to the nearer sub-jet,
$\min_i\bigl||\eta_{\mathrm{sj},i}| - \eta_c\bigr|$, which uses the known position of the step, as a real calibration would use the known geometry of the detector.

All coordinates are supplied unsigned, a simplification permitted by the injected
distortion, which is symmetric about the middle of the detector. The two halves of a real
detector are not identical, and a calibration of real jets would require the signed coordinates, which our method could accommodate.

The response of the recombined jet then depends on the sub-jet axis orientation ${\sjangle}$.  Jets are divided into two classes by the orientation of the sub-jet axis. Those
with ${\sjangle} \approx 0$ or $\pm\pi$ have that axis along the longitudinal
direction and their sub-jets sit at different distances from the step; those with
${\sjangle} \approx \pm\pi/2$ have it along the azimuth and their sub-jets sit at the
same distance. The two classes are drawn separately throughout.

\begin{figure}
    \centering
    \includegraphics[width=0.45\linewidth]{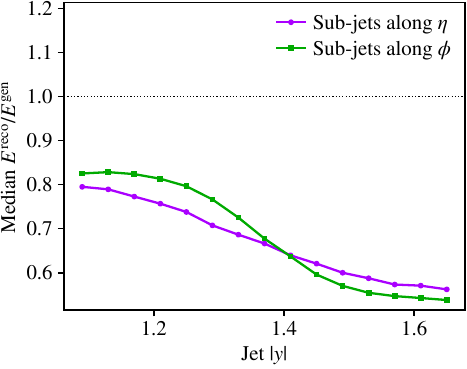}
     \includegraphics[width=0.45\linewidth]{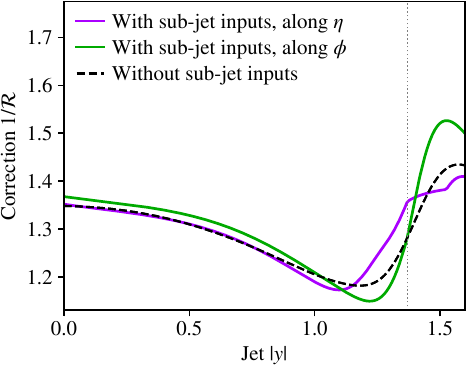}
    \caption{Left, Uncorrected response $E^{\mathrm{reco}}/E^{\mathrm{gen}}$ vs large-$R$-jet
    $|y|$, sliced by sub-jet angle ${\sjangle}$. Right, Network correction $1/\mathcal{R}$ probed at fixed $E$ vs large-$R$-jet
    $|y|$ for the two ${\sjangle}$ configurations; the network with sub-jet inputs (solid) separates them, the network without (black dashed), does not.}
    \label{fig:subjet-uncorr}
\end{figure}

Figure~\ref{fig:subjet-uncorr} shows on the left the uncorrected response against
the large-$R$-jet coordinate. A jet whose sub-jets
lie along the azimuth $\phi$ carries both of them across the transition together, so its
response falls steeply as the jet crosses. A jet whose sub-jets lie along the
longitudinal direction $\eta$ carries them across one at a time, so its response falls
over a range set by their separation. 
The right-hand panel probes what each network learned, by evaluating the
correction $1/\mathcal{R}$ at a fixed energy of 400~GeV as the coordinate is
scanned across the step. The network without the sub-jet inputs returns one curve
for both classes, while the network with the sub-jet inputs returns two.

\begin{figure}
    \centering
    \includegraphics[width=0.95\linewidth]{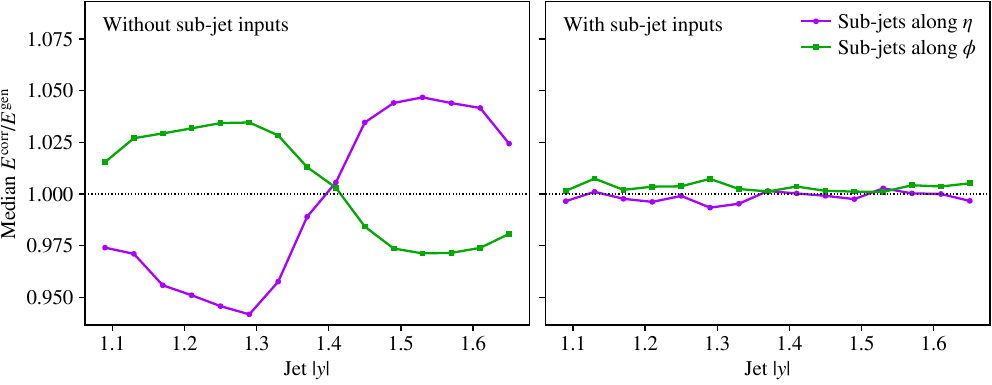}
    \caption{Corrected response vs large-$R$-jet $|y|$, sliced by ${\sjangle}$, for the
    corrections without (left) and with (right) sub-jet inputs.}
    \label{fig:subjet-corr}
\end{figure}

\begin{figure}
    \centering
    \includegraphics[width=0.95\linewidth]{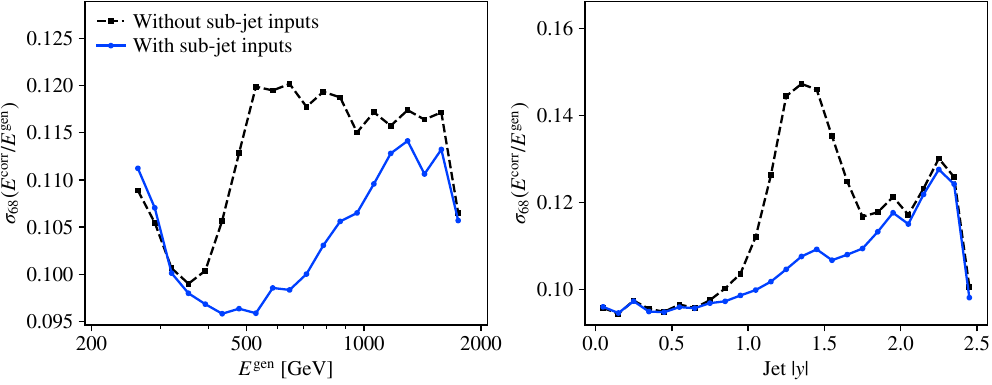}
    \caption{Resolution $\sigma_{68}(E/E^{\mathrm{gen}})$ vs $E^{\mathrm{gen}}$
    (left) and large-$R$-jet $|y|$ (right), with and without sub-jet inputs.}
    \label{fig:subjet-sigma68}
\end{figure}

Figure~\ref{fig:subjet-corr} shows the closure for networks with and without the sub-jet information. With the sub-jet information, the network brings the two classes towards unity.  Without it, it applies the same correction to both, giving a wider spread in response closure. Figure~\ref{fig:subjet-sigma68} shows the resolution against energy and against
the large-$R$-jet coordinate. The improvement is concentrated near the step, which is
where the two classes differ, and elsewhere the two agree. 


\section{Differences between simulation and data}
\label{sec:mcdata}

The corrections of \secref{sec:pileup} and \secref{sec:abs} are derived in
simulation and applied to both simulation and data. Where the simulation
describes the detector imperfectly, applying the same correction to both leaves a
difference between them, which is targeted by the {\it in situ} correction stage.  The correction is measured by balancing jets against a well-measured object, and applied to data.

ATLAS uses several such balances, each covering a different range. A jet in the
central region is balanced against a $Z$ boson or a photon; a jet at higher transverse momentum than any single boson can
reach is balanced against a recoiling system of several jets; and a jet in the
forward region, where no such reference is available, is balanced against a
central jet which has already been calibrated by the first two.~\footnote{A precise measurement of the jet energy scale derived from single-particle measurements has also been employed in combinations with {\it in situ} balance techniques~\cite{ATLAS:2024kkj}.}
Two of these are implemented here; related work has also studied calibration methods that use the correlated information in two jets simultaneously~\cite{SeeingDouble}.

For jets within $|\eta| < 0.8$, the reference used here is the particle-level jet, standing
in for the high precision measurement of the momentum of a $Z$ boson decaying to leptons. The dijet sample already in hand is used for simplicity and consistency, and the substitution idealizes the reference as perfectly measured, where a real $Z$ has imperfect resolution, though nearly negligible compared to the jet energy resolution.  For jets with $|\eta| > 0.8$, the reference is a central jet, and the two are balanced against each other. No multi-jet balance is implemented, nor the missing-transverse-momentum projection fraction method used alongside the direct balance~\cite{CMS:2016lmd,ATLAS:2020cli}, which would be a natural further term of the loss.

\subsection{Injection}

The  sample of simulated $pp\rightarrow jj$ collisions is divided in half, one half labeled data and distorted by a factor $f$, which takes one of five
forms. All five share a constant offset $d_0$ and a slope in the logarithm of the
transverse momentum $d_1$, and they differ in how they depend on pseudorapidity:
\begin{align}
  f_{\mathrm{linear}}    &= 1 + d_0 + d_1\ln(p_{\textrm{T}}/p_{\textrm{T}}^{0}) + d_2|\eta|, \\
  f_{\mathrm{quadratic}} &= 1 + d_0 + d_1\ln(p_{\textrm{T}}/p_{\textrm{T}}^{0}) + d_2|\eta|^2, \\
  f_{\mathrm{threshold}} &= 1 + d_0 + d_1\ln(p_{\textrm{T}}/p_{\textrm{T}}^{0})
                              + d_2\,\mathrm{sign}(|\eta| - 1.5), \\
  f_{\mathrm{expo}}      &= 1 + d_0 + d_1\ln(p_{\textrm{T}}/p_{\textrm{T}}^{0})
                              + d_2\,\frac{e^{|\eta|}-1}{e^{2.5}-1},
\end{align}
with $p_{\textrm{T}}^{0} = 45$~GeV, so that each reduces to $1 + d_0$ for a central jet at
that transverse momentum.  The fifth form adds a dependence on the pileup density $\rho$,
\begin{equation}
  f_{\mathrm{\rho\eta}} = 1 + d_0 + d_1\ln(p_{\textrm{T}}/p_{\textrm{T}}^{0}) + d_2|\eta|^2
    + d_3\left(\frac{\rho}{\rho^{0}} - 1\right)\frac{p_{\textrm{T}}^{0}}{p_{\textrm{T}}},
\end{equation}
with $\rho^{0} = 15$~GeV per unit area, to demonstrate the NN method's capacity for extension to additional observables.

The coefficients are $(d_0, d_1, d_2) = (0.10, 0.08, 0.03)$ for the four
kinematic forms and $(d_0, d_1, d_2, d_3) = (0.10, 0.08, 0.03, 0.05)$ for the
$\rho\eta$ form.

The injected differences change only the scale of the jets, not their resolution, and the particle-level reference idealizes the $Z$ balance. The quantile loss of \secref{sec:mcdata_nn} is built to match the shape of the response as well as its position; testing it on differences in resolution is left to future work.

\subsection{\texorpdfstring{Parametric method}{Parametric method}}
\label{sec:mcdata_atlas}

ATLAS measures a ratio of ratios:
\begin{equation}
  c = \frac{\mathcal{R}^{\mathrm{data}}}{\mathcal{R}^{\mathrm{MC}}},
\end{equation}
to capture the performance in the data relative to the simulation.~\cite{ATLAS:2020cli}. It is derived against the reference transverse momentum,
converted to a function of the jet transverse momentum by the numerical inversion
of \secref{sec:abs_atlas} and applied to data.

ATLAS derives it in three stages: a dijet balance which calibrates forward jets
against central ones, a $Z/\gamma$ balance which calibrates the central region in
absolute terms and a multijet balance which extends beyond the reach of either
boson. The first stage makes the response uniform in
pseudorapidity and a correction derived centrally in the second therefore applies
everywhere.  The first, dijet, stage has no external reference, but measures the relative response of one region to another, generating an overdetermined system of equations.  
 The first two are implemented here.


The correction is {\it derived} as a function of a reference transverse momentum rather
than the measured one. This is necessary, because at
fixed measured transverse momentum, the jets present are preferentially those which
fluctuated upward, due to the steeply falling jet momentum spectrum, and the average response exceeds unity even when data and
simulation agree. This dependence on the conditioning variable and on the underlying spectrum is a general calibration-bias issue~\cite{MLCalibrationBias}. The {\it applied} correction, however, must be a function of the measured transverse momentum.  A jet measured at
$p_{\textrm{T}}^{\mathrm{reco}}$ belongs to a reference of about $p_{\textrm{T}}^{\mathrm{reco}}\,s$, so
the applied scale is the solution of
\begin{equation}
  s = c\bigl(p_{\textrm{T}}^{\mathrm{reco}}\,s,\,|\eta|\bigr),
\end{equation}
which is found by iterating from $s = c(p_{\textrm{T}}^{\mathrm{reco}})$.

\subsection{NN method}
\label{sec:mcdata_nn}

\subsubsection{The scale}

The network produces the scalar $s$, the factor by which the
transverse momentum of a jet in data is scaled. It is a function of three inputs,
the transverse momentum, $|\eta|$ and the pile-up density $\rho$, each
standardized using the mean and width of the simulated training sample. The third
is supplied only for the $f_{\mathrm{\rho\eta}}$ distortion as a demonstration of the network's extendability.

The two stages of \secref{sec:mcdata_atlas} are built into the form of the scale
rather than applied one after the other. It is factorized as
\begin{equation}
  s = s_{\mathrm{abs}}\;s_{\mathrm{rel}},
\end{equation}
an absolute piece which is common to all jets and a relative piece which
describes how the forward region differs from the central one. Each is produced
by a network of two hidden layers of width 64 with rectified linear activations,
$\mathrm{Linear}(\cdot,64) \to \mathrm{ReLU} \to \mathrm{Linear}(64,64) \to
\mathrm{ReLU} \to \mathrm{Linear}(64,1)$. Both
final layers are initialized at zero, so that training begins from $s = 1$.

 The absolute scale is a function of
the transverse momentum alone, to describe an overall difference in scale.
For the distortion which depends on
the pile-up density it is given $\rho$ as well.  The relative scale is constructed so that it is exactly unity in the central
region. Writing $r$ for the network which produces it,
\begin{equation}
  \log s_{\mathrm{rel}} = r\bigl(p_{\textrm{T}},\,\max(|\eta|,\eta_c),\,\rho\bigr)
                        - r\bigl(p_{\textrm{T}},\,\eta_c,\,\rho\bigr),
  \label{eq:srel}
\end{equation}
with $\eta_c = 0.8$ the boundary between the two regions. For a central jet, the
two terms cancel and carry no gradient, fixing the gauge. The central term
of the loss can move only $s_{\mathrm{abs}}$, and the forward term is evaluated
with $s_{\mathrm{abs}}$ held fixed so that it can move only $s_{\mathrm{rel}}$.
Thus the central region anchors the forward one, though the two stages are  separated by construction rather than by being run
in sequence.

\subsubsection{Loss and Training}

The loss has one term for each of the two balances,
$\mathcal{L} = \mathcal{L}_{\mathrm{reference}} + \mathcal{L}_{\mathrm{balance}}$. The first acts on jets with $|\eta| < 0.8$, where as described earlier, the particle-level jet serves the role of the well-measured probe, as no $Z$ boson exists in these samples. In bins of $\eta$ and $p_{\textrm{T}}$, the corrected response
$r = s\,p_{\textrm{T}}^{\mathrm{reco}}/p_{\textrm{T}}^{\mathrm{gen}}$ is evaluated in data and in
simulation, and 19 quantiles spanning the range from $5\%$ to $95\%$ are taken of
each, to match the distribution in shape not only in position.  The loss is the weighted
mean over bins of the squared differences,
\begin{equation}
  \mathcal{L}_{\mathrm{reference}}
    = \frac{\sum_b w_b\,
        \mathrm{MSE}\bigl(Q[r_b^{\mathrm{data}}],\,Q[r_b^{\mathrm{MC}}]\bigr)}
           {\sum_b w_b},
  \qquad w_b = \sqrt{N_b^{\mathrm{data}}},
\end{equation}
with the simulated quantiles held fixed as the target.

The second acts on pairs of jets in which the outer jet lies beyond $|\eta| =
0.8$. In bins of the two pseudorapidities and of the pair-average transverse
momentum, the median logarithmic balance
${b} = \log(s_2 p_{T,2}) - \log(s_1 p_{T,1})$ is taken in data and in
simulation, and the loss is the weighted mean of the squared difference between
them,
\begin{equation}
  \mathcal{L}_{\mathrm{balance}}
    = \frac{\sum_c w_c\,
        \bigl(\mathrm{med}[{b}_c^{\mathrm{data}}]
            - \mathrm{med}[{b}_c^{\mathrm{MC}}]\bigr)^2}{\sum_c w_c},
  \qquad w_c = \sqrt{N_c^{\mathrm{data}}}.
\end{equation}
Pairs with one jet
just beyond the boundary are tied to the central gauge through
Eq.~(\ref{eq:srel}), and pairs of two forward jets are tied in turn to those, so
that the anchor is relayed outward through the chain of bins, acting as stepping stones to fix the far-forward bins relative to the center.

The numerical inversion from reference to measured scales is done as described above for the parametric method.

Optimization uses Adam{~\cite{kingma2017adammethodstochasticoptimization}} with a learning rate of $5 \times 10^{-4}$ and a weight
decay of $10^{-4}$, for 500 epochs in a single phase with no curriculum. The
weights of the final epoch are used.

\subsection{Performance}
\label{sec:mcdata_perf}

In-situ corrections are derived and applied on the output of the NN pipeline of
\secref{sec:pileup} and \secref{sec:abs}, to  isolate this stage. Performance is measured by comparing the distributions of the response ($p_{\textrm{T}}^{corr}/p_{\textrm{T}}^{gen}$) and the dijet asymmetry ($A = (p_{T,1} - p_{T,2})/\overline{p_{\textrm{T}}}$) between data and simulation, via the Kolmogorov--Smirnov distance. Values of $D_{\rm KS}^{r}$ and $D_{\rm KS}^{A}$ are given in Tab.~\ref{tab:mcdata}.

\begin{figure}
    \centering
    \includegraphics[width=0.3\linewidth]{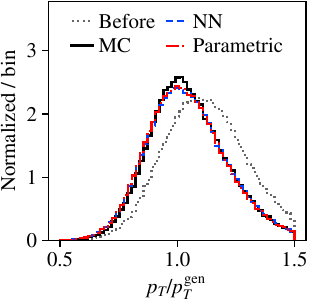}
    \includegraphics[width=0.3\linewidth]{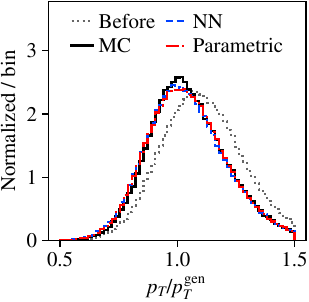}
     \includegraphics[width=0.3\linewidth]{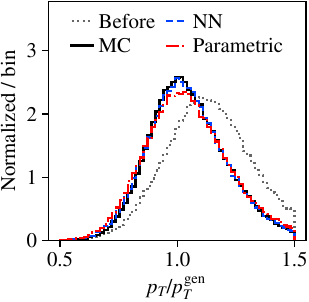}
      \includegraphics[width=0.3\linewidth]{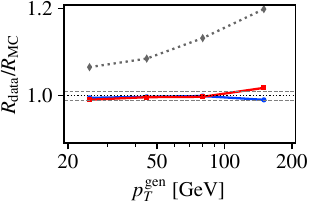}
    \includegraphics[width=0.3\linewidth]{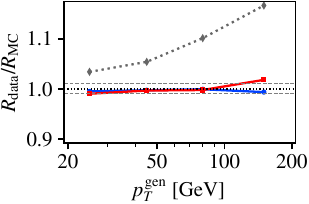}
     \includegraphics[width=0.3\linewidth]{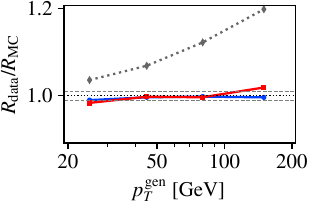}
       \includegraphics[width=0.3\linewidth]{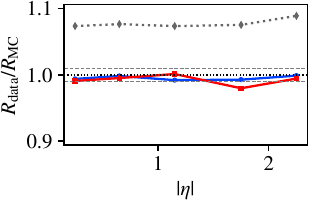}
    \includegraphics[width=0.3\linewidth]{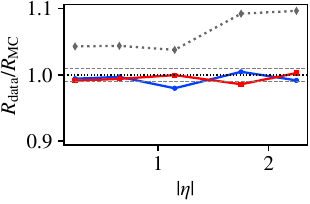}
     \includegraphics[width=0.3\linewidth]{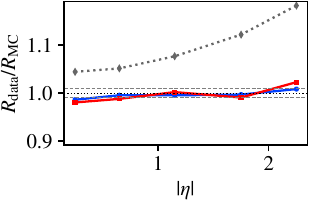}
         \includegraphics[width=0.3\linewidth]{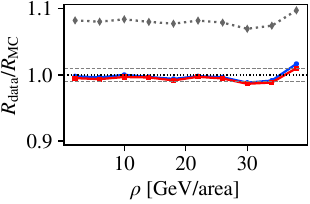}
    \includegraphics[width=0.3\linewidth]{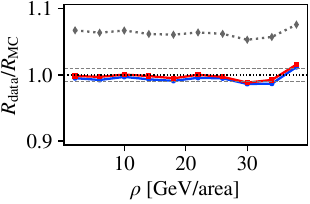}
     \includegraphics[width=0.3\linewidth]{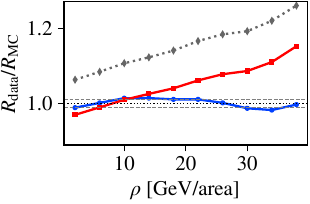}
    \caption{Performance of the in-situ corrections for three distortions: expo, threshold and $\rho\eta$ by columns.  The top row shows the distribution of the response in the data $R=p_{\textrm{T}}^{\textrm{corr}}/p_{\textrm{T}}^{\textrm{gen}}$ before correction, after the NN and parametric corrections, and for the MC target. Rows below show the double ratio of response in data to that in MC as a function of transverse momentum, or pseudorapidity or pileup density $\rho$.}
    \label{fig:insitu_resp}
\end{figure}

\begin{figure}
    \centering
    \includegraphics[width=0.3\linewidth]{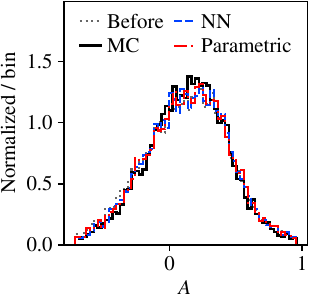}
    \includegraphics[width=0.3\linewidth]{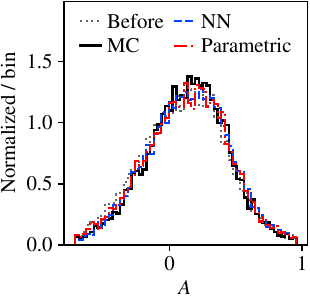}
     \includegraphics[width=0.3\linewidth]{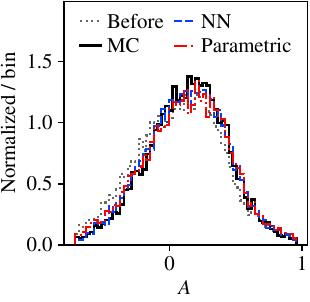}
      \includegraphics[width=0.3\linewidth]{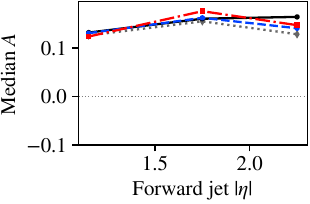}
    \includegraphics[width=0.3\linewidth]{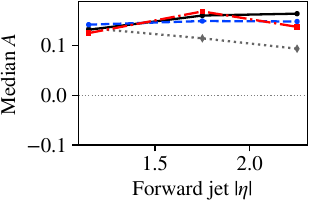}
     \includegraphics[width=0.3\linewidth]{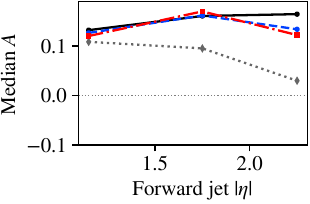}
    \caption{Dijet asymmetry $A$ for the in-situ corrections, for the expo, threshold and $\rho\eta$ distortions by columns. The top row shows the distribution of $A$ in data before correction, after the NN and parametric corrections, and in simulation; the bottom row shows the median of $A$ against the $|\eta|$ of the forward jet.}
    \label{fig:insitu_asym}
\end{figure}

Figure~\ref{fig:insitu_resp} shows the response for three distortions, before and
after correction and against transverse momentum, pseudorapidity and $\rho$.
Figure~\ref{fig:insitu_asym} shows the asymmetry, before and after correction and
against pseudorapidity. The clearest difference, for the $\rho$-dependent distortion, comes from an input, $\rho$, which the functional form of the baseline does not have; it is not a claim that the baseline is poorly tuned.

\begin{table}[]
    \centering
    \caption{Agreement between corrected data and simulation for the NN and parametric in-situ corrections: the Kolmogorov--Smirnov distance between the two distributions of the response ($D_\text{KS}^{r}$) and of the dijet asymmetry ($D_\text{KS}^{A}$), in units of $10^{-3}$; smaller is better. Values are the mean and spread over five divisions of the sample into simulation and data. The row ``nom'' has no distortion.}
\label{tab:mcdata}
\begin{tabular}{lcccc}
\hline\hline
 & \multicolumn{2}{c}{NN (combined-QM)} & \multicolumn{2}{c}{Param.} \\
\cline{2-3} \cline{4-5}
distortion & $D_\text{KS}^{r}$ & $D_\text{KS}^{A}$ & $D_\text{KS}^{r}$ & $D_\text{KS}^{A}$ \\
\hline
nom & 9 $\pm$ 2 & 11 $\pm$ 3 & 5 $\pm$ 4 & 9 $\pm$ 2 \\
linear & 12 $\pm$ 2 & 19 $\pm$ 4 & 14 $\pm$ 4 & 20 $\pm$ 3 \\
quad & 12 $\pm$ 2 & 19 $\pm$ 4 & 19 $\pm$ 4 & 28 $\pm$ 4 \\
threshold & 17 $\pm$ 5 & 16 $\pm$ 2 & 13 $\pm$ 4 & 19 $\pm$ 4 \\
expo & 14 $\pm$ 5 & 18 $\pm$ 2 & 15 $\pm$ 5 & 17 $\pm$ 3 \\
rhoeta & 11 $\pm$ 3 & 19 $\pm$ 4 & 23 $\pm$ 3 & 29 $\pm$ 4 \\
\hline\hline
\end{tabular}
\end{table}

\section{Nuisance-parameter treatment}
\label{sec:np}

In the ATLAS approach, statistical uncertainties on the coefficients of the parametric corrections contribute to the total jet uncertainty, though small compared to the various systematic uncertainties.  The analogous quantities in the NN method are the thousands of network weights. However, uncertainties on the many weights do not need to produce many thousands of new nuisance parameters. The existing ATLAS procedure to decompose a high-dimensional jet-energy-scale nuisance-parameter space into a smaller set of dominant modes~\cite{ATLAS:2014rjc,ATLAS:2020cli} can be applied to capture the statistical uncertainty on the weights in just a few parameters.

The uncertainty of the network's in-situ correction of \secref{sec:mcdata} is measured by repeating the derivation. Data and simulation are
resampled with replacement, event by event, and the correction is
retrained on each of 30 replicas, each from a different random initialization, producing a bundle of corrections.

The bundle is reduced by principal component analysis. The nominal is the
geometric mean over replicas,
$z_0(\mathbf{x}) = \exp(\langle \log s^{(b)}(\mathbf{x})\rangle_b)$, taken in the
logarithm because a scale factor is multiplicative. The residuals
$\delta^{(b)} = \log s^{(b)} - \log z_0$ give a covariance over the grid treated
as a single vector,
\begin{equation}
  C = \frac{1}{B-1}\sum_b \delta^{(b)}\delta^{(b)\mathsf{T}},
\end{equation}
of size $(N_{p_{\textrm{T}}}N_{|\eta|}) \times (N_{p_{\textrm{T}}}N_{|\eta|})$. Its off-diagonal blocks
carry the correlation between transverse momentum and pseudorapidity. Writing the eigendecomposition
$C = \sum_i \lambda_i \phi_i \phi_i^\mathsf{T}$, the correction with its
uncertainty is
\begin{equation}
  s(\mathbf{x}, \boldsymbol{\nu})
    = z_0(\mathbf{x})\,
      \exp\Bigl(\sum_i \nu_i \sqrt{\lambda_i}\,\phi_i(\mathbf{x})\Bigr),
  \label{eq:np}
\end{equation}
with independent unit-Gaussian nuisance parameters $\nu_i$, each moving the
correction along one two-dimensional shape $\phi_i$.

\begin{figure}
    \centering
    \includegraphics[width=0.45\linewidth]{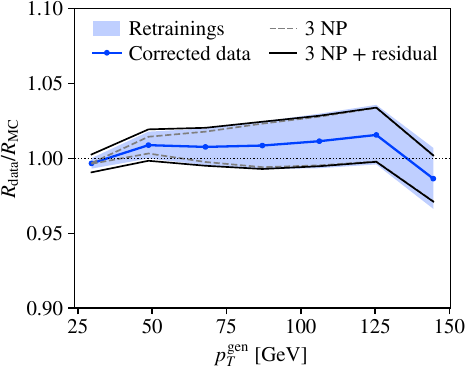}
        \includegraphics[width=0.45\linewidth]{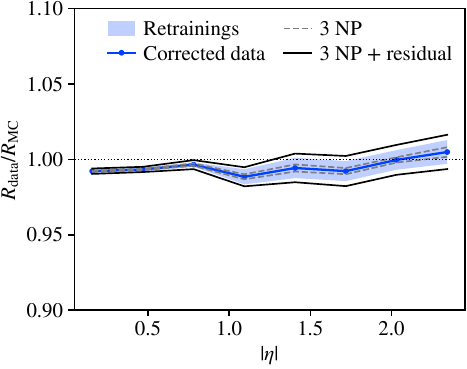}
    \caption{Corrected double ratio versus particle-level \pt\ (left) and $|\eta|$ (right).
The curve is the median of 30 retrainings and the shaded band their $\pm1\sigma$
spread in each bin. The lines are the $\pm1\sigma$ band of Eq.~(\ref{eq:np_res}) with three
parameters (dashed) and with the residual added (solid).}
    \label{fig:np_tests}
\end{figure}

\begin{figure}
    \includegraphics[width=0.45\linewidth]{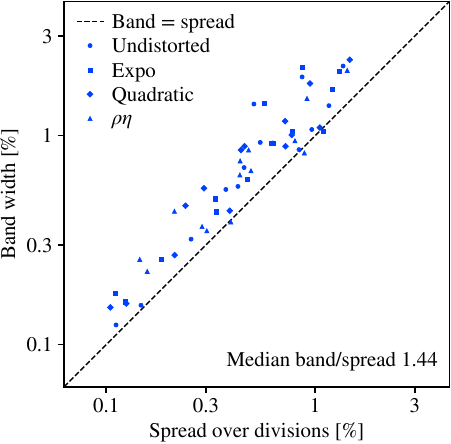}
    \caption{Width of the uncertainty band against the spread observed over 20 divisions of the sample into simulation and data, one point per region of $|\eta|$ and $p_{\textrm{T}}$ and per distortion.}
    \label{fig:np_spread}
\end{figure}

In this application, the leading eigenvalue carries $84\%$ of the variance and
the leading three carry $95\%$.  These fractions are taken over the whole grid, so they are dominated by the region
where the variance is largest, at high transverse momentum.  As in the ATLAS reduction~\cite{ATLAS:2014rjc}, the
remaining components are  combined into one residual parameter, whose shape
is, at each point, the width the three leading components leave out,
\begin{equation}
  \sigma_r^2(\mathbf{x}) = C(\mathbf{x},\mathbf{x}) - \sum_{i\le 3}\lambda_i\,\phi_i^2(\mathbf{x}),
\end{equation}
so that the full expression is
\begin{equation}
  s(\mathbf{x}, \boldsymbol{\nu})
    = z_0(\mathbf{x})\,
      \exp\Bigl(\sum_{i\le 3} \nu_i \sqrt{\lambda_i}\,\phi_i(\mathbf{x})
      + \nu_r\,\sigma_r(\mathbf{x})\Bigr).
  \label{eq:np_res}
\end{equation}
The four parameters reproduce the variance of the bundle exactly at every point, without the correlations. Figure~\ref{fig:np_tests} shows the corrected double ratio
$R_{\mathrm{data}}/R_{\mathrm{MC}}$ with two bands and the spread of the retrainings, and the band produced by varying the reduced nuisance parameters. 
Figure~\ref{fig:np_spread} compares the width of the uncertainty band to the observed spread for several distortions.

\section{Discussion}
\label{sec:discussion}

The pipeline replaces each stage of the jet calibration with a neural network trained directly against the quantity that stage corrects. No functional form is imposed on the correction, so extending it means adding inputs rather than engineering new forms: the pile-up network absorbs dependence on transverse momentum, non-linear terms and an interaction between the pile-up variables (\secref{sec:pileup} and the Appendix), and the large-radius calibration gains the positions of the sub-jets (\secref{sec:subjet_y}), in each case without a change of method.

At nearly every stage, the networks match or exceed the parametric baseline. The residual pile-up dependence is on average about a tenth of the baseline's (Table~\ref{tab:pu}). The non-closure of the absolute scale over the grid of energy and pseudorapidity is about half the baseline's in the lo-$p_\textrm{T}$ regime and a quarter in the hi-$p_\textrm{T}$ regime (Table~\ref{tab:abs_closure}); closure is defined there by the median, which the network targets, and with the peak of the core instead the baseline closes slightly better in the lo-$p_\textrm{T}$ regime. For the differences between data and simulation (Table~\ref{tab:mcdata}), the two methods agree within the spread over divisions of the sample for the undistorted sample and for the linear, exponential and step distortions, and the network is better for the quadratic and $\rho$-dependent ones.

The statistical uncertainty of a network correction need not be carried by its thousands of weights. For the in-situ correction, retraining on resampled data and simulation and decomposing the resulting corrections into principal components reduces it to a few nuisance parameters, each a correlated shape in transverse momentum and pseudorapidity (\secref{sec:np}), as is done for the conventional uncertainties. This study is a prototype: it uses a fast detector simulation with injected distortions, and the systematic uncertainties of the in-situ methods are not treated; they would be propagated by retraining the correction for each variation and reducing the resulting corrections with the same decomposition. Natural extensions are to merge the stages into a single network, to add further jet features as inputs, and to correct the jet direction as well as its energy. A natural next step is to replace \textsc{Delphes} with a learned simulation such as Parnassus~\cite{Parnassus}.

\appendix

\section{Additional pileup distortions}
\subsection{Mis-estimated forward pile-up density}
\label{sec:etadens}

The $|\eta|$ dependence injected in \secref{sec:pileup} is imposed directly through the ad hoc factor $(1 + a_2|\eta|)$. An
alternative is described here in which the residual dependence is instead a
consequence of the pile-up density being mis-estimated.

The density $\rho$ is estimated in ATLAS from $k_t$ jets built from
positive-energy calorimeter clusters within $|\eta| \le 2$, and the measured
value falls to nearly zero beyond $|\eta| \simeq 2$~\cite{ATLAS:2015ull}, where
calorimeter cells are coarser. Clusters are
seeded on the significance of a cell energy relative to its noise~\cite{ATLASTopoClusters}, so a coarser cell
is less likely to be seeded by the soft deposits pile-up leaves. Forward pile-up
is therefore under-represented in the median, and $\rho \times A$
under-subtracts forward jets. What survives is a dependence on the pile-up
conditions which grows with $|\eta|$. In ATLAS, this leftover is absorbed by the
residual correction, whose coefficients are fit in bins of
$|\eta_{\mathrm{det}}|$.

Here, an excess density is added before the subtraction, as
\begin{equation}
  p_{\textrm{T}}^{\rm reco} \;\to\; p_{\textrm{T}}^{\rm reco} + \rho\,A\,\big(f(|\eta|)-1\big),
  \qquad
  f(|\eta|) - 1 \;=\; d\,\sigma\!\left(\frac{|\eta|-\eta_c}{c_w}\right),
\end{equation}
with $\sigma$ the logistic function, $d$ the fractional under-report far forward,
$\eta_c$ the position of the transition and $c_w$ its width. The parameters are a depth $d = 0.5$, so that the true forward density is
$1.5$ times the reported $\rho$, a transition at $\eta_c = 2.0$ following
Ref.~\cite{ATLAS:2015ull} and a width $c_w = 0.15$. 

Only the excess
$(f-1)$ is added, since the \textsc{Delphes} pile-up is already present in
$p_{\textrm{T}}^{\rm reco}$ and already described by the event $\rho$. Subtracting
$\rho \times A$ then removes the part of the density the estimate describes and
leaves
\begin{equation}
  r(|\eta|) \;=\; \rho\,A\; d\,\sigma\!\left(\frac{|\eta|-\eta_c}{c_w}\right),
  \label{eq:etadens_resid}
\end{equation}
which vanishes in the central region and grows forward. The value of $\rho$ used
in the subtraction is unchanged, so the residual arises  from the
mismatch between the density present and the density reported.

\begin{figure}
    \centering
    \includegraphics[width=0.95\linewidth]{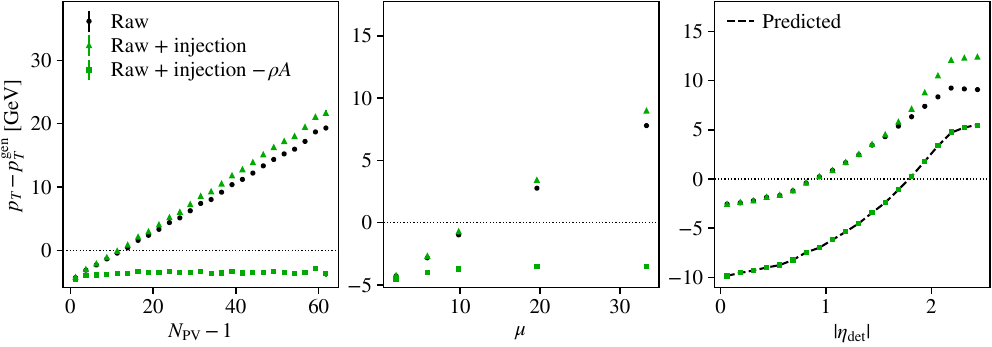}    \caption{Forward pile-up density study: the residual $p_{\textrm{T}}-p_{\textrm{T}}^{\rm gen}$ against $N_{\rm PV}-1$ (left), $\mu$ (center) and $|\eta_{\rm det}|$ (right), before the injection, after it, and after the injection followed by the $\rho\times A$ subtraction. The dashed line is the expectation of Eq.~(\ref{eq:etadens_resid}).}
    \label{fig:etadens_validation}
\end{figure}

No $\alpha$ or $\beta$ dependence is injected, and the absolute residual described below is not included to avoid swamping the effect of interest here. No mock $\mu'$ is constructed: $\mu$ is the single value assigned to each simulated sample, and so carries nearly the same information as $N_{\rm PV}$.
Figure~\ref{fig:etadens_validation} shows the residual
$p_{\textrm{T}}^{\rm reco} - p_{\textrm{T}}^{\rm gen}$ against $N_{\rm PV}-1$, against $\mu$ and
against $|\eta|$, before the injection, after it, and after the injection
followed by the $\rho \times A$ subtraction. 

\begin{figure}[ht]
  \centering
  \includegraphics[width=0.85\linewidth]{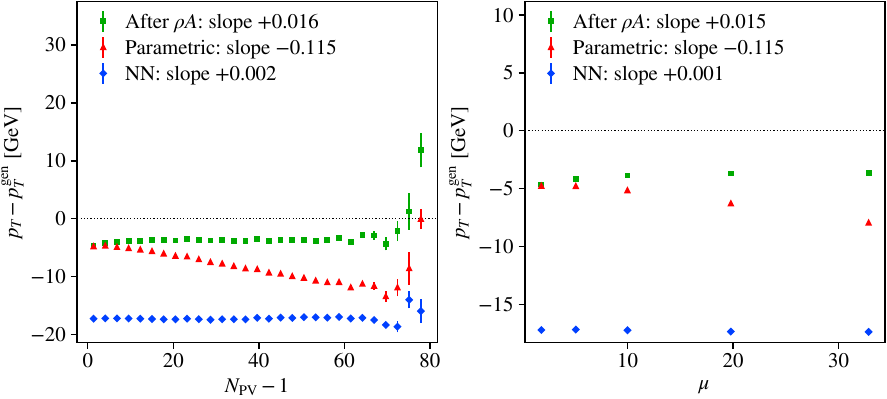}
        \includegraphics[width=0.85\linewidth]{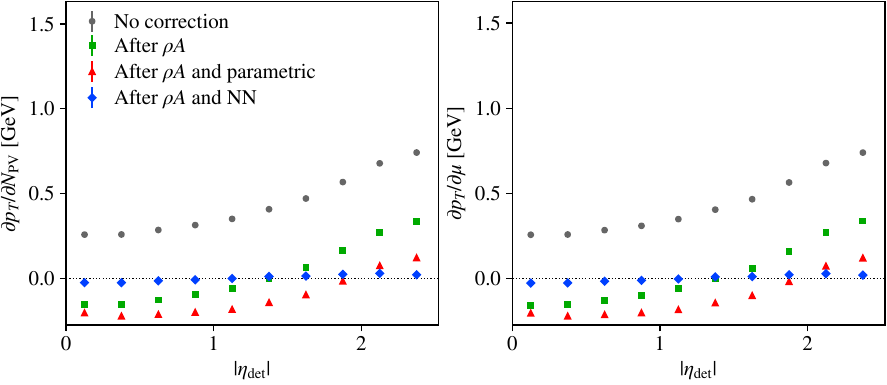}
  \caption{Performance of the NN-based and parametric pile-up corrections for the
    $\eta$-density injection: residual versus $N_{\rm PV}$ and $\mu$ before and
    after correction, and $\partial p_{\textrm{T}}/\partial N_{\rm PV}$ and
    $\partial p_{\textrm{T}}/\partial \mu$ in bins of $|\eta_{\rm det}|$.}
  \label{fig:etadens_perf}
\end{figure}

The figures of merit are those of \secref{sec:pileup_perf}, the root mean square
across bins of $|\eta_{\mathrm{det}}|$ of the post-correction slope of the
residual against each pile-up variable.  Table~\ref{tab:etadens} gives the result and Figure~\ref{fig:etadens_perf} shows
the correction for this dataset.  The last step of the parametric baseline fits the
coefficients to a straight line in $|\eta_{\mathrm{det}}|$, which
cannot describe the non-linear transition injected here. The
network takes the pseudorapidity as an input and is under no such constraint. The gain comes from a dependence outside the functional form of the baseline, not from a poorly tuned baseline.

\begin{table}[ht]
  \centering
   \caption{Residual pile-up dependence after PU residual correction for the $\eta$-density
    injection, for the NN and parametric methods. Entries are the RMS across the
    $|\eta|$ bins of Fig.~\ref{fig:etadens_perf} of the post-correction slope of $p_{\textrm{T}}$
    against each pile-up variable, in GeV per unit $N_{\rm PV}$ (resp.\ $\mu$); zero is ideal.
    The row ``$\rho A$ only'' is the injected sample with only $\rho \times A$ removed, shown
    also in Fig.~\ref{fig:etadens_perf}. The two columns of each pair nearly coincide because
    $\mu$, a single value per simulated sample, carries nearly the same information as
    $N_{\rm PV}$ here.}
  \label{tab:etadens}
\begin{tabular}{l cc cc}
\hline \hline
 & \multicolumn{2}{c}{all $|\eta|$ bins} & \multicolumn{2}{c}{$|\eta| \geq 2$ only} \\
\cmidrule(lr){2-3} \cmidrule(lr){4-5}
 & $\partial p_{\textrm{T}}/\partial N_{\mathrm{PV}}$ & $\partial p_{\textrm{T}}/\partial \mu$ & $\partial p_{\textrm{T}}/\partial N_{\mathrm{PV}}$ & $\partial p_{\textrm{T}}/\partial \mu$ \\
 \hline 
$\rho A$ only    & 0.172 & 0.172 & 0.306 & 0.306 \\
NN               & 0.019 & 0.019 & 0.025 & 0.024 \\
Param.           & 0.162 & 0.162 & 0.099 & 0.099 \\
\midrule
NN/Param.        & 0.12 & 0.12 & 0.25 & 0.25 \\
\hline \hline
\end{tabular}
  
\end{table}

\subsection{Non-linear pile-up dependence}
\label{sec:pu_nonlinear}

\begin{figure}[h!]
    \centering
    \includegraphics[width=0.95\linewidth]{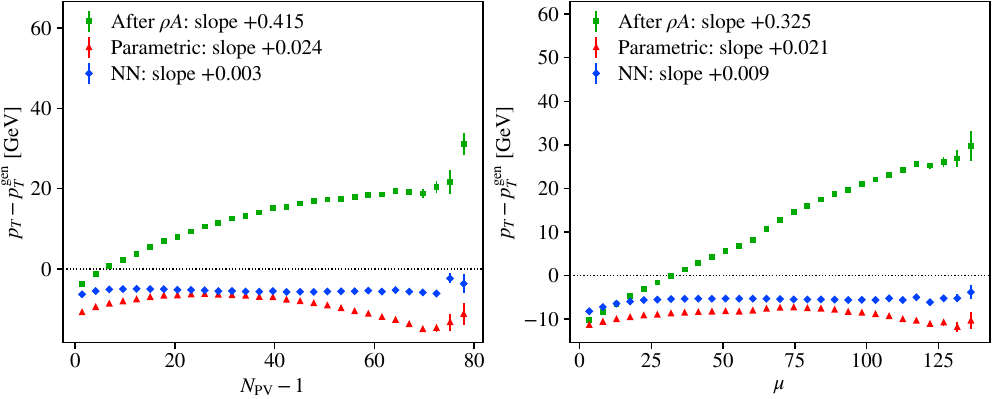}
    \caption{Residual between reconstructed and true jet transverse momenta versus the number of primary vertices (left) and the mean number of PU interactions (right). Shown are means before correction, after NN correction and after parametric correction.}
    \label{fig:pu_nonlin_resid}
\end{figure}

The parametric correction, Equation~(\ref{eq:atlas_pu}), assumes a linear dependence on $N_{\rm PV}$ and in $\mu$. Here, we inject a non-linear dependence, adding a  quadratic term in the
number of primary vertices,
\begin{equation}
  p_{\textrm{T}}^{\rm reco} \;\to\; p_{\textrm{T}}^{\rm reco}
    + \alpha(p_{\textrm{T}}^{\rm gen},|\eta|)\,(N_{\rm PV}-1)
    + \beta (p_{\textrm{T}}^{\rm gen},|\eta|)\,\mu
    + \gamma(p_{\textrm{T}}^{\rm gen},|\eta|)\,(N_{\rm PV}-1)^2 ,
  \label{eq:pu_nonlin}
\end{equation}
with $\gamma$ carrying the same dependence on the jet as the linear
coefficients,
\begin{equation}
  \gamma(p_{\textrm{T}},|\eta|) = (g_0 + g_1 \ln p_{\textrm{T}})\,(1 + g_2 |\eta|),
\end{equation}
and $g_0 = -0.0022$, $g_1 = -0.0010$, $g_2 = 0.15$. The samples are otherwise those of \secref{sec:pileup}, in the lo-$p_{\textrm{T}}$
regime, and three values of $\rho_{\mu N}$ are selected.   Figure~\ref{fig:pu_nonlin_resid} shows the residual against $N_{\rm PV}$ and
against $\mu$ for $\rho_{\mu N} = 0.5$, before and after corrections.




\subsection{Interaction between the pile-up variables}
\label{sec:pu_cross}

The parametric correction, Equation~(\ref{eq:atlas_pu}), includes no terms which are a product of  $N_{\rm PV}$ and $\mu$.

Ref.~\cite{ATLAS:2015ull}  anticipates cross-terms with shorter bunch spacing, where  the in-time and
out-of-time contributions overlap in the same calorimeter cells.

Here, we inject  such a term,
\begin{equation}
  p_{\textrm{T}}^{\rm reco} \;\to\; p_{\textrm{T}}^{\rm reco}
    + \alpha(p_{\textrm{T}}^{\rm gen},|\eta|)\,(N_{\rm PV}-1)
    + \beta (p_{\textrm{T}}^{\rm gen},|\eta|)\,\mu
    + \kappa\,(N_{\rm PV}-1)\,(\mu - \langle\mu\rangle) ,
  \label{eq:pu_cross}
\end{equation}
with $\kappa = 0.010$~GeV. The term vanishes on average, since $\mu$ enters
through its deviation from the sample mean, so it shifts no scale and adds only a
correlation: at fixed $N_{\rm PV}$ the response depends on how much out-of-time
pile-up accompanied it.

\begin{figure}[h!]
    \centering
    \includegraphics[width=0.95\linewidth]{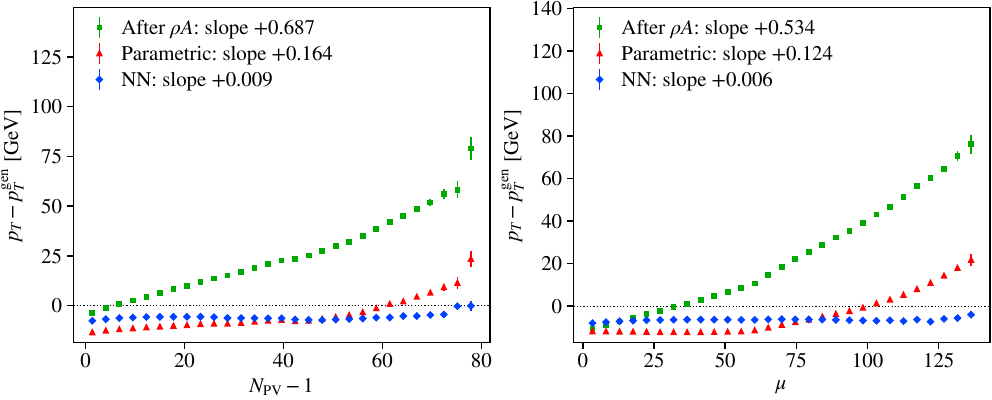}
    \caption{Residual between reconstructed and true jet transverse momenta versus the number of primary vertices (left) and the mean number of PU interactions (right), for the interaction injection of Eq.~(\ref{eq:pu_cross}) with $\rho_{\mu N}=0.5$. Shown are means before correction, after NN correction and after parametric correction.}
    \label{fig:pu_cross_resid}
\end{figure}

Several values of $\rho_{\mu N}$ are selected. At $\rho_{\mu N} = 0$ the two pile-up variables are independent
and the product is a direction which neither variable spans alone. At
$\rho_{\mu N} = 1$ they are proportional, the product degenerates into a
quadratic in $N_{\rm PV}$. Figure~\ref{fig:pu_cross_resid} shows the residual against $N_{\rm PV}$ and
against $\mu$ for $\rho_{\mu N} = 0.5$, before correction and after each of the
two corrections.



\section*{Data availability}
The simulated samples used in this article are openly available~\cite{jetcorrml_data}.
The code that produces every figure and table of this article from these samples is
openly available~\cite{jetcorrml_code}; its repository is
\url{https://github.com/danielwhiteson/jetcorrml}.

\begin{acknowledgments}
B.N. is supported by the U.S. Department of Energy (DOE), Office of Science under contract DE-AC02-76SF00515. D.W. is supported by the DOE Office of Science. BM gratefully acknowledges the continuous support from LPNHE, CNRS/IN2P3, Sorbonne Universit\'e and Universit\'e de Paris Cit\'e. SS acknowledges that this work is part of a project that has received funding from the European Research Council (ERC) under the European Union’s Horizon 2020 research and innovation programme (Grant agreement No. 948254) and from the Swiss National Science Foundation (SNSF) under Eccellenza grant number PCEFP2\_194658.
\end{acknowledgments}

\bibliography{mljes_param}

@article{ATLAS:2019oxp,
    author = "{ATLAS Collaboration}",
    title = "{Determination of jet calibration and energy resolution in proton-proton collisions at $\sqrt{s}$ = 8 TeV using the ATLAS detector}",
    eprint = "1910.04482",
    archivePrefix = "arXiv",
    primaryClass = "hep-ex",
    reportNumber = "CERN-EP-2019-057",
    doi = "10.1140/epjc/s10052-020-08477-8",
    journal = "Eur. Phys. J. C",
    volume = "80",
    number = "12",
    pages = "1104",
    year = "2020"
}

@article{ATLAS:2020cli,
    author = "{ATLAS Collaboration}",
    title = "{Jet energy scale and resolution measured in proton{\textendash}proton collisions at $\sqrt{s}=13$~TeV with the ATLAS detector}",
    eprint = "2007.02645",
    archivePrefix = "arXiv",
    primaryClass = "hep-ex",
    reportNumber = "CERN-EP-2020-083",
    doi = "10.1140/epjc/s10052-021-09402-3",
    journal = "Eur. Phys. J. C",
    volume = "81",
    number = "8",
    pages = "689",
    year = "2021"
}

@article{ATLAS:2018bip,
    author = "{ATLAS Collaboration}",
    title = "{In situ calibration of large-radius jet energy and mass in 13 TeV proton{\textendash}proton collisions with the ATLAS detector}",
    eprint = "1807.09477",
    archivePrefix = "arXiv",
    primaryClass = "hep-ex",
    reportNumber = "CERN-EP-2018-191",
    doi = "10.1140/epjc/s10052-019-6632-8",
    journal = "Eur. Phys. J. C",
    volume = "79",
    number = "2",
    pages = "135",
    year = "2019"
}

@article{ATLAS:2015ull,
    author = "{ATLAS Collaboration}",
    title = "{Performance of pile-up mitigation techniques for jets in $pp$ collisions at $\sqrt{s}=8$  TeV using the ATLAS detector}",
    eprint = "1510.03823",
    archivePrefix = "arXiv",
    primaryClass = "hep-ex",
    reportNumber = "CERN-PH-EP-2015-206",
    doi = "10.1140/epjc/s10052-016-4395-z",
    journal = "Eur. Phys. J. C",
    volume = "76",
    number = "11",
    pages = "581",
    year = "2016"
}

@article{ATLAS:2024kkj,
    author = "{ATLAS Collaboration}",
    title = "{A precise measurement of the jet energy scale derived from single-particle measurements and in situ techniques in proton-proton collisions at $\sqrt{s}=$ 13 TeV with the ATLAS detector}",
    eprint = "2407.15627",
    archivePrefix = "arXiv",
    primaryClass = "hep-ex",
    reportNumber = "CERN-EP-2024-195",
    doi = "10.1140/epjc/s10052-025-14409-1",
    journal = "Eur. Phys. J. C",
    volume = "85",
    number = "9",
    pages = "927",
    year = "2025"
}

@article{CMS:2011shu,
    author = "{CMS Collaboration}",
    title = "{Determination of Jet Energy Calibration and Transverse Momentum Resolution in CMS}",
    eprint = "1107.4277",
    archivePrefix = "arXiv",
    primaryClass = "physics.ins-det",
    reportNumber = "CERN-PH-EP-2011-102, CMS-JME-10-011",
    doi = "10.1088/1748-0221/6/11/P11002",
    journal = "JINST",
    volume = "6",
    pages = "P11002",
    year = "2011"
}

@article{Agarwal:2022txa,
    author = "Agarwal, Garvita",
    collaboration = "CMS",
    title = "{Jet Energy Scale and Resolution Measurements in CMS}",
    eprint = "2301.02175",
    archivePrefix = "arXiv",
    primaryClass = "hep-ex",
    doi = "10.22323/1.414.0652",
    journal = "PoS",
    volume = "ICHEP2022",
    pages = "652",
    year = "2022"
}

@article{ATLAS:2018kot,
  author        = "{ATLAS Collaboration}",
  title         = "{Observation of $H \rightarrow b\bar{b}$ decays and $VH$ production with the ATLAS detector}",
  journal       = "Phys. Lett. B",
  volume        = "786",
  pages         = "59--86",
  year          = "2018",
  doi           = "10.1016/j.physletb.2018.09.013",
  eprint        = "1808.08238",
  archivePrefix = "arXiv",
  primaryClass  = "hep-ex",
}

@article{CMS:2018nsn,
  author        = "{CMS Collaboration}",
  title         = "{Observation of Higgs Boson Decay to Bottom Quarks}",
  journal       = "Phys. Rev. Lett.",
  volume        = "121",
  pages         = "121801",
  year          = "2018",
  doi           = "10.1103/PhysRevLett.121.121801",
  eprint        = "1808.08242",
  archivePrefix = "arXiv",
  primaryClass  = "hep-ex",
}

@article{ATLAS:2019fgd,
  author        = "{ATLAS Collaboration}",
  title         = "{Search for new resonances in mass distributions of jet pairs using 139 fb$^{-1}$ of $pp$ collisions at $\sqrt{s}=13$ TeV with the ATLAS detector}",
  journal       = "JHEP",
  volume        = "03",
  pages         = "145",
  year          = "2020",
  doi           = "10.1007/JHEP03(2020)145",
  eprint        = "1910.08447",
  archivePrefix = "arXiv",
  primaryClass  = "hep-ex",
}

@article{CMS:2019gwf,
  author        = "{CMS Collaboration}",
  title         = "{Search for high mass dijet resonances with a new background prediction method in proton-proton collisions at $\sqrt{s}=13$ TeV}",
  journal       = "JHEP",
  volume        = "05",
  pages         = "033",
  year          = "2020",
  doi           = "10.1007/JHEP05(2020)033",
  eprint        = "1911.03947",
  archivePrefix = "arXiv",
  primaryClass  = "hep-ex",
}

@article{ATLAS:2014rjc,
  author        = "{ATLAS Collaboration}",
  title         = "{Jet energy measurement and its systematic uncertainty in proton--proton collisions at $\sqrt{s}=7$ TeV with the ATLAS detector}",
  journal       = "Eur. Phys. J. C",
  volume        = "75",
  pages         = "17",
  year          = "2015",
  doi           = "10.1140/epjc/s10052-014-3190-y",
  eprint        = "1406.0076",
  archivePrefix = "arXiv",
  primaryClass  = "hep-ex",
}

@article{CMS:2016lmd,
  author        = "{CMS Collaboration}",
  title         = "{Jet energy scale and resolution in the CMS experiment in $pp$ collisions at 8 TeV}",
  journal       = "JINST",
  volume        = "12",
  pages         = "P02014",
  year          = "2017",
  doi           = "10.1088/1748-0221/12/02/P02014",
  eprint        = "1607.03663",
  archivePrefix = "arXiv",
  primaryClass  = "hep-ex",
}

@article{Guest:2018yhq,
  author        = "Guest, Dan and Cranmer, Kyle and Whiteson, Daniel",
  title         = "{Deep Learning and its Application to LHC Physics}",
  journal       = "Ann. Rev. Nucl. Part. Sci.",
  volume        = "68",
  pages         = "161--181",
  year          = "2018",
  doi           = "10.1146/annurev-nucl-101917-021019",
  eprint        = "1806.11484",
  archivePrefix = "arXiv",
  primaryClass  = "hep-ex",
}

@techreport{ATLAS:2020nnc,
  author      = "{ATLAS Collaboration}",
  title       = "{Simultaneous Jet Energy and Mass Calibrations with Neural Networks}",
  institution = "CERN",
  address     = "Geneva",
  number      = "ATL-PHYS-PUB-2020-001",
  year        = "2020",
  url         = "https://cds.cern.ch/record/2706189",
}

@article{CMS:2019bjr,
  author        = "{CMS Collaboration}",
  title         = "{A Deep Neural Network for Simultaneous Estimation of $b$ Jet Energy and Resolution}",
  journal       = "Comput. Softw. Big Sci.",
  volume        = "4",
  pages         = "10",
  year          = "2020",
  doi           = "10.1007/s41781-020-00041-z",
  eprint        = "1912.06046",
  archivePrefix = "arXiv",
  primaryClass  = "hep-ex",
}

@article{ATLAS:2023dnn,
  author        = "{ATLAS Collaboration}",
  title         = "{Simultaneous energy and mass calibration of large-radius jets with the ATLAS detector using a deep neural network}",
  journal       = "Mach. Learn. Sci. Tech.",
  year          = "2024",
  doi           = "10.1088/2632-2153/ad611e",
  eprint        = "2311.08885",
  archivePrefix = "arXiv",
  primaryClass  = "hep-ex",
}

@article{ATLAS:2023hbp,
  author        = "{ATLAS Collaboration}",
  title         = "{New techniques for jet calibration with the ATLAS detector}",
  journal       = "Eur. Phys. J. C",
  volume        = "83",
  pages         = "761",
  year          = "2023",
  doi           = "10.1140/epjc/s10052-023-11837-9",
  eprint        = "2303.17312",
  archivePrefix = "arXiv",
  primaryClass  = "hep-ex",
}

@techreport{ATLAS:2018gni,
  author      = "{ATLAS Collaboration}",
  title       = "{Generalized Numerical Inversion: A Neural Network Approach to Jet Calibration}",
  institution = "CERN",
  address     = "Geneva",
  number      = "ATL-PHYS-PUB-2018-013",
  year        = "2018",
  url         = "https://cds.cern.ch/record/2630972",
}

@article{Cukierman:2016ntt,
  author        = "Cukierman, Aviv and Nachman, Benjamin",
  title         = "{Mathematical Properties of Numerical Inversion for Jet Calibrations}",
  journal       = "Nucl. Instrum. Meth. A",
  volume        = "858",
  pages         = "1--11",
  year          = "2017",
  doi           = "10.1016/j.nima.2017.03.038",
  eprint        = "1609.05195",
  archivePrefix = "arXiv",
  primaryClass  = "physics.data-an",
}

@article{madgraph,
	Archiveprefix = {arXiv},
	Author = {Alwall, J. and Frederix, R. and Frixione, S. and Hirschi, V. and Maltoni, F. and Mattelaer, O. and Shao, H. -S. and Stelzer, T. and Torrielli, P. and Zaro, M.},
	Doi = {10.1007/JHEP07(2014)079},
	Eprint = {1405.0301},
	Journal = {JHEP},
	Pages = {079},
	Primaryclass = {hep-ph},
	Reportnumber = {CERN-PH-TH-2014-064, CP3-14-18, LPN14-066, MCNET-14-09, ZU-TH-14-14},
	Slaccitation = {%%CITATION = ARXIV:1405.0301;%%},
	Title = {{The automated computation of tree-level and next-to-leading order differential cross sections, and their matching to parton shower simulations}},
	Volume = {07},
	Year = {2014}}

@article{delphes,
	Archiveprefix = {arXiv},
	Author = {de Favereau, J. and others},
	Collaboration = {DELPHES 3},
	Doi = {10.1007/JHEP02(2014)057},
	Eprint = {1307.6346},
	Journal = {JHEP},
	Pages = {057},
	Primaryclass = {hep-ex},
	Slaccitation = {%%CITATION = ARXIV:1307.6346;%%},
	Title = {{DELPHES 3, A modular framework for fast simulation of a generic collider experiment}},
	Volume = {1402},
	Year = {2014}}

@article{Cacciari:2008gp,
    author = "Cacciari, Matteo and Salam, Gavin P. and Soyez, Gregory",
    title = "{The anti-$k_t$ jet clustering algorithm}",
    eprint = "0802.1189",
    archivePrefix = "arXiv",
    primaryClass = "hep-ph",
    reportNumber = "LPTHE-07-03",
    doi = "10.1088/1126-6708/2008/04/063",
    journal = "JHEP",
    volume = "04",
    pages = "063",
    year = "2008"
}

@article{Gallicchio:2018elx,
    author = "Gallicchio, Jason and Chien, Yang-Ting",
    title = "{Quit Using Pseudorapidity, Transverse Energy, and Massless Constituents}",
    eprint = "1802.05356",
    archivePrefix = "arXiv",
    primaryClass = "hep-ph",
    month = "2",
    year = "2018"
}

@misc{kingma2017adammethodstochasticoptimization,
      title={Adam: A Method for Stochastic Optimization}, 
      author={Diederik P. Kingma and Jimmy Ba},
      year={2017},
      eprint={1412.6980},
      archivePrefix={arXiv},
      primaryClass={cs.LG},
      url={https://arxiv.org/abs/1412.6980}, 
}

@article{Larkoski:2014wba,
  author        = "Larkoski, Andrew J. and Marzani, Simone and Soyez, Gregory and Thaler, Jesse",
  title         = "{Soft Drop}",
  journal       = "JHEP",
  volume        = "05",
  pages         = "146",
  year          = "2014",
  doi           = "10.1007/JHEP05(2014)146",
  eprint        = "1402.2657",
  archivePrefix = "arXiv",
  primaryClass  = "hep-ph",
}

@article{Dasgupta:2013ihk,
  author        = "Dasgupta, Mrinal and Fregoso, Alessandro and Marzani, Simone and Salam, Gavin P.",
  title         = "{Towards an understanding of jet substructure}",
  journal       = "JHEP",
  volume        = "09",
  pages         = "029",
  year          = "2013",
  doi           = "10.1007/JHEP09(2013)029",
  eprint        = "1307.0007",
  archivePrefix = "arXiv",
  primaryClass  = "hep-ph",
}

@article{CacciariSalamPileup,
    author = "Cacciari, Matteo and Salam, Gavin P.",
    title = "{Pileup subtraction using jet areas}",
    eprint = "0707.1378",
    archivePrefix = "arXiv",
    primaryClass = "hep-ph",
    reportNumber = "LPTHE-07-01",
    doi = "10.1016/j.physletb.2007.09.077",
    journal = "Phys. Lett. B",
    volume = "659",
    pages = "119--126",
    year = "2008"
}

@article{Pythia8,
    author = {Sj{\"o}strand, Torbj{\"o}rn and Ask, Stefan and Christiansen, Jesper R. and Corke, Richard and Desai, Nishita and Ilten, Philip and Mrenna, Stephen and Prestel, Stefan and Rasmussen, Christine O. and Skands, Peter Z.},
    title = "{An introduction to PYTHIA 8.2}",
    eprint = "1410.3012",
    archivePrefix = "arXiv",
    primaryClass = "hep-ph",
    reportNumber = "LU-TP-14-36, MCNET-14-22, CERN-PH-TH-2014-190, FERMILAB-PUB-14-316-CD, DESY-14-178, SLAC-PUB-16122",
    doi = "10.1016/j.cpc.2015.01.024",
    journal = "Comput. Phys. Commun.",
    volume = "191",
    pages = "159--177",
    year = "2015"
}

@article{ATLASTopoClusters,
    author = "Aad, Georges and others",
    collaboration = "ATLAS",
    title = "{Topological cell clustering in the ATLAS calorimeters and its performance in LHC Run 1}",
    eprint = "1603.02934",
    archivePrefix = "arXiv",
    primaryClass = "hep-ex",
    reportNumber = "CERN-PH-EP-2015-304",
    doi = "10.1140/epjc/s10052-017-5004-5",
    journal = "Eur. Phys. J. C",
    volume = "77",
    pages = "490",
    year = "2017"
}

@article{DetectorResponseNN,
    author = "Cheong, Sanha and Cukierman, Aviv and Nachman, Benjamin and Safdari, Murtaza and Schwartzman, Ariel",
    title = "{Parametrizing the Detector Response with Neural Networks}",
    eprint = "1910.03773",
    archivePrefix = "arXiv",
    primaryClass = "physics.data-an",
    doi = "10.1088/1748-0221/15/01/P01030",
    journal = "JINST",
    volume = "15",
    number = "01",
    pages = "P01030",
    year = "2020"
}

@article{MLCalibrationBias,
    author = "Gambhir, Rikab and Nachman, Benjamin and Thaler, Jesse",
    title = "{Bias and priors in machine learning calibrations for high energy physics}",
    eprint = "2205.05084",
    archivePrefix = "arXiv",
    primaryClass = "hep-ph",
    reportNumber = "MIT-CTP 5432",
    doi = "10.1103/PhysRevD.106.036011",
    journal = "Phys. Rev. D",
    volume = "106",
    number = "3",
    pages = "036011",
    year = "2022"
}

@article{SeeingDouble,
    author = "Gambhir, Rikab and Nachman, Benjamin",
    title = "{Calibrating two jets at once}",
    eprint = "2402.14067",
    archivePrefix = "arXiv",
    primaryClass = "hep-ph",
    reportNumber = "MIT-CTP 5680",
    doi = "10.1103/PhysRevD.110.076015",
    journal = "Phys. Rev. D",
    volume = "110",
    number = "7",
    pages = "076015",
    year = "2024"
}

@article{PUMML,
    author = "Komiske, Patrick T. and Metodiev, Eric M. and Nachman, Benjamin and Schwartz, Matthew D.",
    title = "{Pileup Mitigation with Machine Learning (PUMML)}",
    eprint = "1707.08600",
    archivePrefix = "arXiv",
    primaryClass = "hep-ph",
    reportNumber = "MIT-CTP-4924",
    doi = "10.1007/JHEP12(2017)051",
    journal = "JHEP",
    volume = "12",
    pages = "051",
    year = "2017"
}

@article{PileupGNN,
    author = "Arjona Mart{\'\i}nez, J. and Cerri, Olmo and Pierini, Maurizio and Spiropulu, Maria and Vlimant, Jean-Roch",
    title = "{Pileup mitigation at the Large Hadron Collider with graph neural networks}",
    eprint = "1810.07988",
    archivePrefix = "arXiv",
    primaryClass = "hep-ph",
    doi = "10.1140/epjp/i2019-12710-3",
    journal = "Eur. Phys. J. Plus",
    volume = "134",
    number = "7",
    pages = "333",
    year = "2019"
}

@inproceedings{AdamW,
    author = "Loshchilov, Ilya and Hutter, Frank",
    title = "{Decoupled Weight Decay Regularization}",
    eprint = "1711.05101",
    archivePrefix = "arXiv",
    primaryClass = "cs.LG",
    month = "11",
    year = "2017"
}

@inproceedings{SGDR,
    author = "Loshchilov, Ilya and Hutter, Frank",
    title = "{SGDR: Stochastic Gradient Descent with Warm Restarts}",
    eprint = "1608.03983",
    archivePrefix = "arXiv",
    primaryClass = "cs.LG",
    month = "8",
    year = "2016"
}

@article{FrequentistCalibration,
  author        = "Gambhir, Rikab and Nachman, Benjamin and Thaler, Jesse",
  title         = "{Learning Uncertainties the Frequentist Way: Calibration and Correlation in High Energy Physics}",
  journal       = "Phys. Rev. Lett.",
  volume        = "129",
  pages         = "082001",
  year          = "2022",
  doi           = "10.1103/PhysRevLett.129.082001",
  eprint        = "2205.03413",
  archivePrefix = "arXiv",
  primaryClass  = "hep-ph",
}

@article{CaloFlowCalibration,
  author        = "Du, Haoxing and Krause, Claudius and Mikuni, Vinicius and Nachman, Benjamin and Pang, Ian and Shih, David",
  title         = "{Unifying Simulation and Inference with Normalizing Flows}",
  journal       = "Phys. Rev. D",
  volume        = "111",
  pages         = "076004",
  year          = "2025",
  eprint        = "2404.18992",
  archivePrefix = "arXiv",
  primaryClass  = "hep-ph",
}

@article{Parnassus,
  author        = "Dreyer, Etienne and Gross, Eilam and Kobylianskii, Dmitrii and Mikuni, Vinicius and Nachman, Benjamin and Soybelman, Nathalie",
  title         = "{Parnassus: An Automated Approach to Accurate, Precise, and Fast Detector Simulation and Reconstruction}",
  journal       = "Phys. Rev. Lett.",
  volume        = "133",
  pages         = "211902",
  year          = "2024",
  doi           = "10.1103/PhysRevLett.133.211902",
  eprint        = "2406.01620",
  archivePrefix = "arXiv",
  primaryClass  = "physics.data-an",
}

@misc{jetcorrml_data,
  author       = {Whiteson, Daniel},
  title        = {{Simulated dijet and large-radius jet samples for ``Learning to Calibrate Jet Energies''}},
  howpublished = {Zenodo, version 1.0},
  year         = {2026},
  note         = {\href{https://doi.org/10.5281/zenodo.24043234}{doi:10.5281/zenodo.24043234}}
}

@misc{jetcorrml_code,
  author       = {Whiteson, Daniel},
  title        = {{jetcorrml: the code of ``Learning to Calibrate Jet Energies''}},
  howpublished = {Zenodo, version v1.0},
  year         = {2026},
  note         = {\href{https://doi.org/10.5281/zenodo.24044717}{doi:10.5281/zenodo.24044717}}
}
\end{document}